\documentclass[a4paper,11pt]{article}
\usepackage{jcappub}
\usepackage{graphicx,epsfig,natbib,color,times,wasysym}

\def \<{\langle}
\def \>{\rangle}

\def\prd{Phys. Rev. D}

\def\apj{Astrophys. J.}
\def\apjl{Astrophys. J. Lett.}
\def\apjs{Astrophys. J.Suppl.}
\def\mnras{Mon. Not. R. Astr. Soc.}

\def\aj{Astr. J.}

\def\jcap{JCAP}

\begin{document}
\title{Large-scale anomalies of the CMB in the curvaton scenario}
\author{Hao Liu\footnote{Also: Institute of High Energy Physics, China Academy of Science, Beijing, China}, Anne Mette Frejsel and Pavel Naselsky}
\affiliation{Niels Bohr Institute \& Discovery Center, Blegdamsvej 17, DK-2100 Copenhagen, Denmark}




\emailAdd{liuhao@nbi.dk}
\emailAdd{frejsel@nbi.dk}
\emailAdd{naselsky@nbi.dk}

\abstract{We extend the curvaton scenario presented by Erickcek et al.~\cite{Erickcek.0808.1570,Erickcek.0907.0705}, to explain how the even-odd multipole asymmetry of the Cosmic Microwave Background (CMB) (also called parity asymmetry,~\citep{Parity1,Parity2}) and power anisotropies can be generated by the curvaton field, which acts as an extra component to the spectrum of adiabatic perturbations in the inflationary epoch. Our work provides a possible cosmic explanation to the CMB large-scale asymmetry problems besides systematics and unknown residuals.}

\maketitle
\flushbottom

\section{Introduction}
\label{sec:into}
The Cosmological Principle states that the Universe should be isotropic and homogeneous on scales above 100 Mpc, which is widely accepted as a basic principle of most cosmological scenarios. This principle can be experimentally tested by galaxy surveys and CMB observations. The SDSS experiment result indicates that the galaxy distribution becomes isotropic and homogeneous at large scales~\citep{sdss1,sdss2,sdss3,sdss4,sdss5,sdss6,sdss7}, which supports the Cosmological Principle well.

However, there are still open questions from the CMB observations--so-called anomalies. Among them is the non-Gaussian cold spot~\citep{Cruz2005MNRAS.356...29C,Cruz2007ApJ...655...11C,2007Sci...318.1612C}, the missing power in the quadrupole in all WMAP releases~\citep{WMAP1,WMAP3,WMAP5,WMAP7,WMAP9} (see however~\citep{WMAP9}), even-odd multipole power asymmetry (also called parity asymmetry~\citep{Parity1,Parity2}), alignment of multipole components and axis-of-evil~\citep{alignment1,alignment2,alignment3,WMAP7 Anomalies discuss}, north-south asymmetry~\citep{NS1,NS2,NS3,NS4,NS5} and so on. Certainly, there are attempts in the listed literatures and others to attribute those anomalies to systematics or unrealized Galactic/Ecliptic emission, or simply the cosmic variance. The systematics can be reduced or eliminated by cross check between independent CMB observation experiments, like WMAP and Planck. As for the cosmic variance, it could explain many anomalies but only with very low probability.
There is an even more fundamental question: whether these anomalies have a common origin or are statistically independent. If we believe that all the anomalies originate from the same source, it would be even more important to discover its origin. In the opposite case, where all the anomalies are statistically independent, the problem is how one peculiar realization of the random field can contain all these anomalies. Although the anomalies have been observed in the temperature data, they could also prove to be main sources of contamination in polarization data. Therefore, the understanding of their origin is potentially crucial for investigating the E-mode and B-mode or even the primordial non-Gaussianities and gravitational waves from inflation.

Therefore, if we believe that the anomalies of the CMB discussed in the literature indicate some constraint on the Cosmological Principle, we should find an explanation through cosmological theory. In the standard inflationary scenario, the large-scale structure is generated by the initial perturbations due to quantum fluctuations of the inflation field. However, if we further consider the possibility that the standard inflation field is not the only field in the inflation stage (the inflation field is still dominating), then by adding some extra components to the nearly scale-invariant spectrum we can introduce a seed of asymmetry to the theoretical expectation, not just to a specific realization by a particular observer, which is the main purpose of this work.

In the curvaton scenario~\citep{Gordon.0607432,Linde.0511736}, we can see that additional non-isotropic perturbations can be generated by the curvaton field, and consequently cause even-odd multipole power asymmetry in the power spectrum. In the curvaton scenario, the curvaton field ($\sigma$) is supposed to have negligible energy density compared to the inflation field. It is also non-interacting with the inflation field, and thus its initial value $\sigma_*$ is kept during inflation, and its quantum fluctuation $(\delta \sigma)_{rms}=H_{inf}/(2\pi)$ ($H_{inf}$ is the Hubble parameter during inflation) contributes part or all of the primordial perturbations~\citep{1990PhRvD..42..313M,1997PhRvD..56..535L,2001PhLB..522..215M,2002PhLB..524....5L}. If the curvaton potential is $V(\sigma)=(1/2)m_\sigma^2\sigma^2$ with $m_\sigma \ll H_{inf}$ ($m_\sigma$ is the mass of the curvaton), then after inflation (where $m_\sigma \simeq H$) the curvaton will oscillate and decay into radiation and will interact with matter. The sequence of curvaton decaying and decoupling of particle species gives different curvaton interacting scenarios, like curvaton-dark matter interacting~\citep{Erickcek.0907.0705}.

The curvaton scenario discussed in this work is an extension of~\cite{Erickcek.0808.1570,Erickcek.0907.0705}. Whereas Erickcek et al. focus on super horizon perturbations (the wave length of the perturbation they consider is very large), we have discovered that if the wave length of the curvaton perturbation is comparable to or smaller than the horizon, then the model can be used for explanation of some of the CMB anomalies.

The outline of this paper is the following. We present the extended model in section \ref{sec:model}. In section \ref{sec:apply to WMAP}, we apply this model to the WMAP data to see if it can, at least partly, explain the power spectrum parity asymmetry and the temperature space anomalies. In section 4 we show how a plane wave component can affect the CMB power spectrum. In the end, a brief discussion is given in section \ref{sec:conclusion}.

\section{Extended model based on the curvaton scenario}
\label{sec:model}
Based on the curvaton scenario, we have constructed a model with only three parameters to see if it can generate some of the observed anomalous features of the CMB, in particular power asymmetry in the power spectrum. The model is presented below.

Following to~\cite{Parity2}, if there is a primordial perturbation in Fourier space $\Psi(\mathbf k)$, then the low multipole ($2\le l\le 30$) spherical
harmonic decomposition coefficients ($a_{lm}$) are connected to $\Psi(\mathbf k)$ through
\begin{equation}\label{equ:alm from prim pert}
a_{lm} = 4\pi (-\imath)^l \int \frac{d^3\mathbf k}{(2\pi)^3} \Psi(\mathbf k)\,T_l(k)\,Y^*_{lm}(\hat {\mathbf k}),
\end{equation}
where $T_l(k)$ is the radiation transfer function. For odd multipoles, $l=2n+1$ ($n=0,1,2,..$),
\begin{equation}\label{equ:odd alm}
a_{lm} = -\frac{(-\imath)^{l-1}}{\pi^2} \int\limits^{\infty}_0 dkk^2 \int\limits^{\pi}_0 d\theta_{\mathbf k} \sin\theta_{\mathbf k}
\times\int\limits^{\pi}_0 d\phi_{\mathbf k} T_l(k) Y^*_{lm}(\hat{\mathbf k}) \mathrm{Im}[\Psi(\mathbf k)],
\end{equation}
and for the even multipoles, $l=2n$,
\begin{equation}\label{equ:even alm}
a_{lm}=\frac{(-\imath)^l}{\pi^2} \int\limits^{\infty}_0 dkk^2 \int\limits^{\pi}_0 d\theta_{\mathbf k} \sin\theta_{\mathbf k}\times\int\limits^{\pi}_0 d\phi_{\mathbf k} T_l(k) Y^*_{lm}(\hat{\mathbf k}) \mathrm{Re}[\Psi(\mathbf k)].
\end{equation}
From Eq.(\ref{equ:odd alm}) and (\ref{equ:even alm}), we can see that e.g. odd-parity preference might be produced, provided that
\begin{equation}
|\mathrm{Re} [\Psi(\mathbf k)]|\ll|\mathrm{Im} [\Psi(\mathbf k)]|\;\;\;(k\lesssim 22/\eta_0),
\label{equ:primordial_odd}
\end{equation}
where $\eta_0$ is the present conformal time. As is seen from~\ref{equ:primordial_odd} the phases of metric perturbations ($\xi = \arctan\left[\frac{\mathrm{Im}(\Psi(\mathbf{k}))}{\mathrm{Re}(\Psi(\mathbf{k}))}\right]$) have to be localized in the vicinity of $\xi\propto\pi/2;3\pi/2$, at least for the range $k\lesssim 20/\eta_0$ to $30/\eta_0$. This shows the possibility of generating even-odd parity asymmetry from specific primordial perturbations.

The squeezed space of phases indicates that for spatial scales $x\gtrsim 4$ Gpc~\cite{Parity2} the homogeneity and isotropy of the perturbations is abnormal. Namely, parity arguments of the CMB leads to the parity asymmetry of the metric perturbations, $\Psi (\vec{x})\approx -\Psi(-\vec{x})$. Let us assume that the origin of those anomalies can be associated with unusual properties of the curvaton field \cite{Erickcek.0808.1570,Erickcek.0907.0705,Parity1}. The potential perturbation at decoupling due to a curvaton field perturbation is given in~\cite{Erickcek.0808.1570}, using a real space form $ \Psi(\tau_{dec},\vec{x})$, as:
\begin{equation}\label{}
    \Psi(\tau_{dec},\vec{x})\simeq -\frac{R}{5}\left [\frac{\Psi(\tau_{dec},\vec{x})}{\frac{9}{10} \Psi_P}\right ]\left [2\left (\frac{\delta_\sigma}{\overline{\sigma}}\right )+\left (\frac{\delta_\sigma}{\overline{\sigma}}\right )^2\right ],
\end{equation}
where $\overline{\sigma}$ is the homogeneous curvaton background, and $\sigma(\vec{x})=\overline{\sigma}+\delta \sigma({\vec{x}})$. $R\equiv\rho_{\sigma}/\rho_{total}$ is the fraction of curvaton in the total energy density just before curvaton decay. The curvaton decay is assumed to be early enough so that $\Psi_P\simeq-(2R/9)\delta\rho_\sigma/\rho_\sigma$.

We set the time-dependent coefficient of $[2(\frac{\delta_\sigma}{\overline{\sigma}})+(\frac{\delta_\sigma}{\overline{\sigma}})^2]$ as $\psi(\tau)$ and suppose $\frac{\delta_\sigma}{\overline{\sigma}}=r\sin(\vec{k}\cdot\vec{x}+\delta)$ (sinusoidal fluctuation for curvaton perturbation, see also Sec. 4 of~\cite{Erickcek.0907.0705}, and $\delta$ is the phase), then we have the spatial distribution of the real space potential as:

\begin{equation}\label{equ:psi-tau-x}
\Psi(\tau,\vec{x}) = \psi(\tau)[2r\sin(\vec{k}\cdot\vec{x}+\delta)+r^2\sin^2(\vec{k}\cdot\vec{x}+\delta)].
\end{equation}

The low order CMB power spectrum consist of Sachs-Wolfe (SW) and Integrated Sachs-Wolfe (ISW) effects. According to ~\cite{Erickcek.0808.1570} the induced SW effect is $\left [\frac{\Delta T}{T}(\hat{n})\right ]_{SW}=\Psi(\tau_{dec})/3$. Therefore, we only have to calculate the ISW effect. According to Equation 16 of~\cite{Erickcek.0808.1570}, the ISW effect is given by:
\begin{equation}\label{}
    \left [\frac{\Delta T}{T}(\hat{n})\right ]_{ISW} = 2\int^{1}_{a_{dec}}\frac{d\Psi\{a,H_0^{-1}[\chi_0-\chi(a)]\hat{n}\}}{da}da ,
\end{equation}
where $\chi(a)\equiv H_0[\tau(a)-\tau_{dec}]$ and $\chi_0\equiv\chi(a=1)=H_0x_{dec}$. If we assume that $r$ is constant this gives:

\begin{equation}\label{}
    \left [\frac{\Delta T}{T}(\hat{n})\right ]_{ISW} = 2[2r\sin(\vec{k}\cdot\vec{x}_{dec}+\delta) + r^2\sin^2(\vec{k}\cdot\vec{x}_{dec}+\delta)] \Psi(a)\mid^1_{a_{dec}}.
\end{equation}

Combining the SW and ISW effect, and letting $\vec{k}\cdot\vec{x}_{dec}=q\pi[1-\cos(\theta)]=q\omega$ ($q$ is the wave number, $\theta$ is the polar coordinate, and we choose the system of coordinates so that $\vec{k}$ is oriented along -$\vec{Z}$), we have gotten the CMB fluctuations due to curvaton perturbations as:
\begin{equation}\label{equ:CMB due to curvaton}
\left [\frac{\Delta T}{T}(\hat{n})\right ] = 4r\Psi_c[\sin(q\omega+\delta)+\frac{r}{2}\sin^2(q\omega+\delta)],
\end{equation}
where $\Psi_c = \Psi(a)\mid^1_{a_{dec}} + \Psi(\tau_{dec})/3$ is a constant and is related only to the overall amplitude.

Now it is clear that in our plane-wave model, the structure of CMB fluctuations due to curvaton perturbations are determined by only three parameters: $q$, $r$ and $\delta$. At the current stage, we consider only the structural term of Equation~\ref{equ:CMB due to curvaton}:
\begin{equation}\label{equ:model final form}
\left [\frac{\Delta T}{T}(\hat{n})\right ]\propto\sin(q\omega+\delta)+\frac{r}{2}\sin^2(q\omega+\delta),
\end{equation}
and let the amplitude from equation \ref{equ:CMB due to curvaton} be a free parameter. From this equation we can get the curvaton component power spectrum.

Equation~\ref{equ:model final form} can also help us understand why the curvaton-based perturbations are so different to the ordinary adiabatic perturbations. The curvaton-based perturbations are proportional to a linear combination of $\sin(q\omega+\delta)$ and $\sin(q\omega+\delta)^2$. Such $sin$-functions have intrinsic power spectrum odd-even parity asymmetry, and, since there are both first and second orders of the $sin$-function in the combination, different parity asymmetry patterns can be easily produced according to their ratio. Moreover, the linear combination in Equation~\ref{equ:model final form} is rotationally symmetric around the wave vector $\vec{k}$ of the sinusoidal perturbation, which provides an axis of rotation symmetry along $\vec{k}$. Globally speaking, such an axis due to the curvaton scenario can easily be a potential source of asymmetry and/or anomaly, even if the exact direction of the axis can not be predicted by the curvaton scenario alone.

\section{Implementation of the curvaton model}
\label{sec:apply to WMAP}
The curvaton model can now be implemented and compared to CMB data. The model is determined by three parameters: the wave number $q$, the curvaton fluctuation strength $r$ and the initial phase $\delta$. Note that changing $\delta$ is very similar to choosing a special spatial position of a particular observer. Firstly, in section \ref{sec:fitwmapspec} we determine the model parameters by fitting the WMAP power spectrum. Then in section \ref{sec:axes} we proceed to find the most optimal orientation of the model based on the WMAP data.

\subsection{Determining model parameters by fitting the WMAP power spectrum}
\label{sec:fitwmapspec}
We apply the model to the WMAP CMB power spectrum. The best fit $\Lambda$CDM power spectrum does not have power asymmetry, but the observed WMAP CMB power spectrum does. Thus, assuming that the WMAP CMB power spectrum is a combination of $\Lambda$CDM and an extra component due to the curvaton field, the power spectrum of this extra component ($C_{extra}$) should display power asymmetry. Therefore we fit our model to the difference between the $\Lambda$CDM best fit power spectrum and the observed WMAP CMB power spectrum.

\begin{figure}
\begin{center}
  \includegraphics[width=0.48\textwidth]{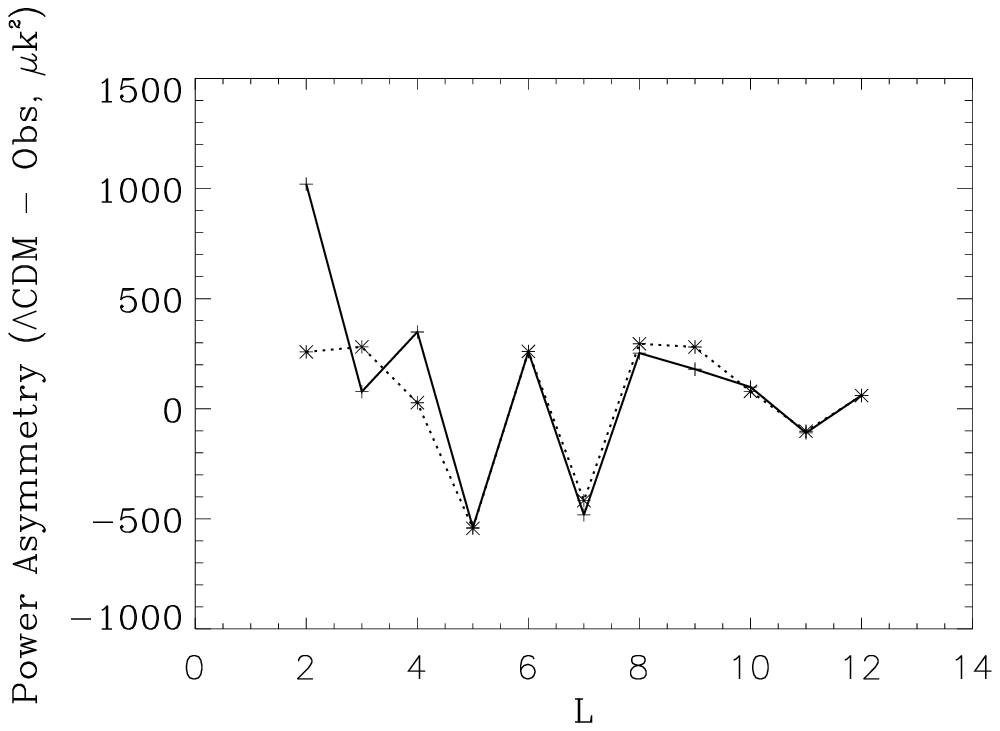}
  \includegraphics[width=0.48\textwidth]{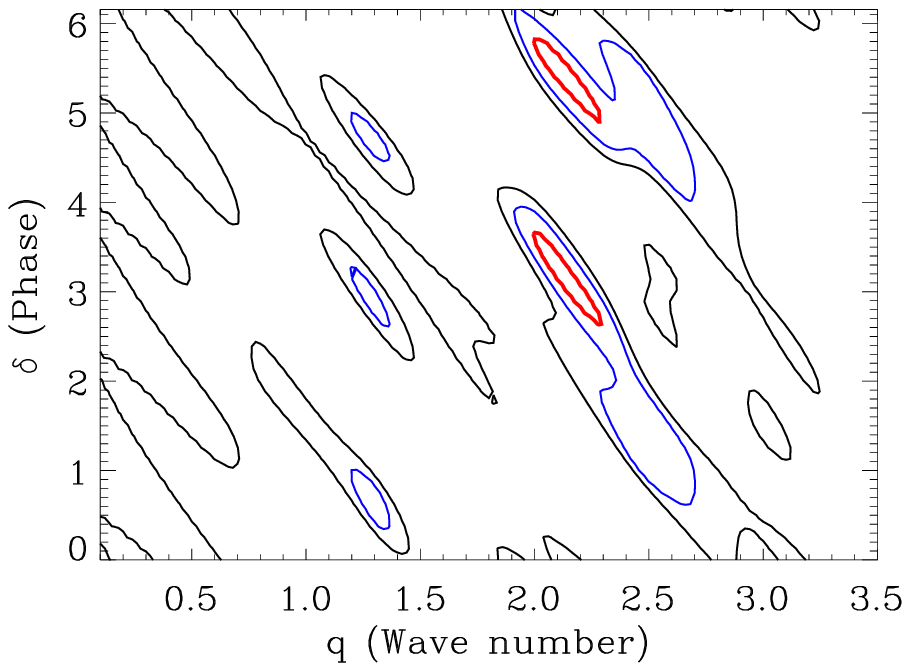}
  \includegraphics[width=0.48\textwidth]{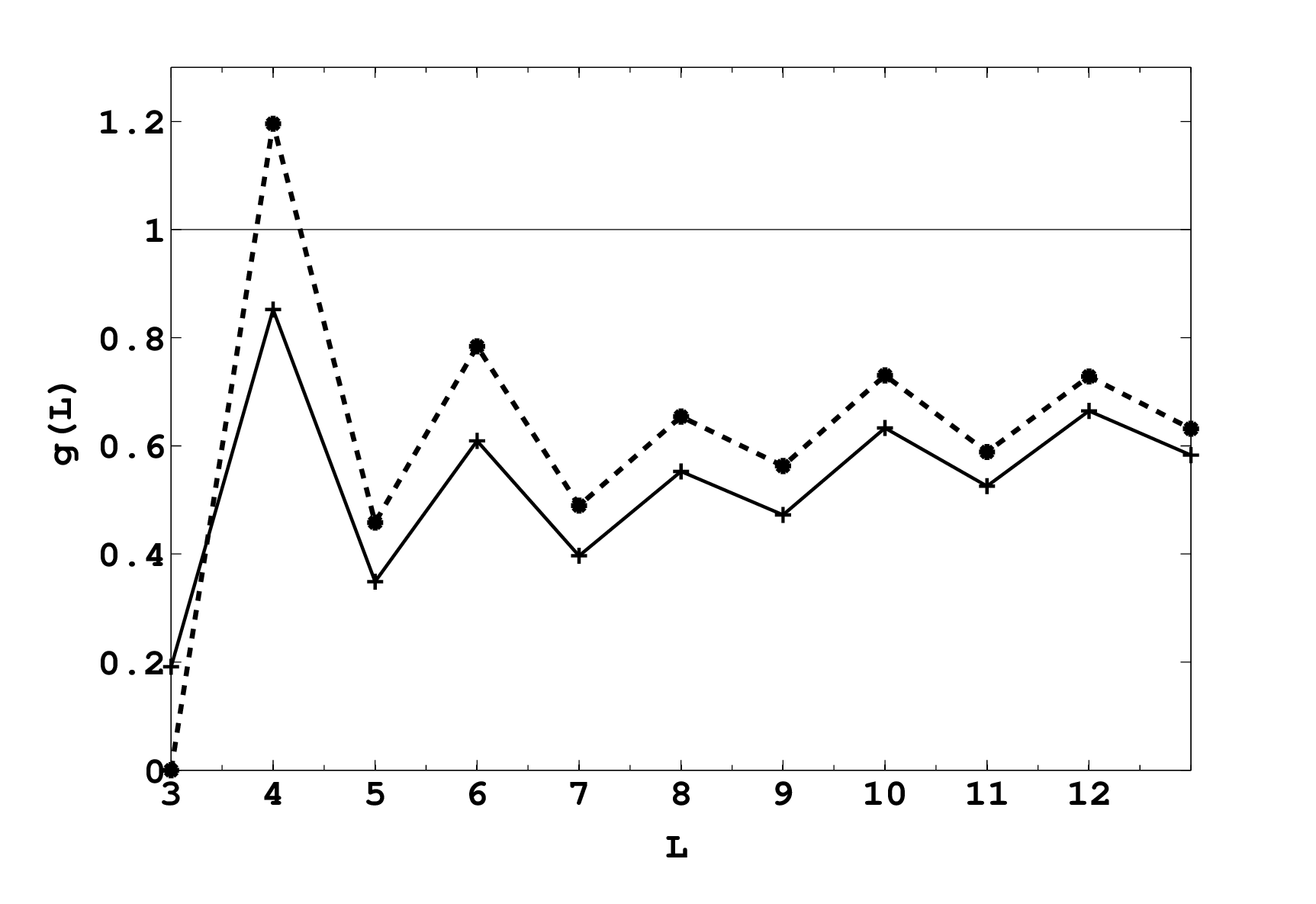}
  \caption{Application of our model to WMAP data: \emph{Top left}: The best fit model power spectrum (dotted) and the difference between the $\Lambda$CDM best fit power spectrum and the observed WMAP 7-yr power spectrum (solid). \emph{Top right}: A 3-level contour plot of $\ln(\mathcal{L}_{fit})$ at 0.5, 0.75, 0.96 (black, blue, red), where $\ln(\mathcal{L}_{fit})$ is normalized to $(0,1)$. \emph{Bottom}: Parity parameter for observed WMAP 7-yr power spectrum (solid line) and for the curvaton model (dashed line).}
\label{fig:fit to WMAP}
\end{center}
\end{figure}

With each parameter set $(q, r, \delta)$, we calculate the CMB temperature distribution according to Equation~\ref{equ:model final form} as well as the CMB power spectrum, and then linearly fit the derived power spectrum to $C_{extra}$ to determine the constant $\Psi_c$ (Equation~\ref{equ:CMB due to curvaton}). The $\chi^2$ statistic of fitting is recorded for this parameter set, and the best guess of $(q,r,\delta)$ is determined by the minimal $\chi^2$. The resulting CMB power spectrum is given in the top left of Fig.~\ref{fig:fit to WMAP}. It seems as though the characteristic power asymmetry structure has been faithfully produced by the model, and only the $l=2$ (quadrupole) component is not particularly well fitted.
To quantitatively estimate whether parity asymmetry is actually produced, we use an estimator $g(l)$ (see \cite{Parity1,Parity2}). The estimator is defined as:
\begin{equation}
	g(l) = \frac{\sum_{l=3}^{l_{max}}l(l+1)C^+_l}{\sum_{l=3}^{l_{max}}l(l+1)C^-_l},
\end{equation}
where $C^+(l) = C(l) \cos^2(\frac{\pi l}{2})$ (the power spectrum for all even $l$) and $C^-(l) = C(l) \sin^2(\frac{\pi l}{2})$ (the power spectrum for all odd $l$). Note that we take the sum from $l=3$ because of the poor fit of the quadrupole. In Fig.~\ref{fig:fit to WMAP} (bottom) we see that the parity asymmetry ($g(l)<1$) is indeed reproduced for the model for low multipoles, as expected.

We also give a contour plot of the likelihood of fitting in the parameter space. The likelihood of fitting can be calculated as $\mathcal{L}_{fit}=P(\chi^2>\chi^2_{fit})$. We have tested that the model CMB power spectrum is not sensitive to $r$ and thus we choose to fix $r$ at its best guess value, $r=2.6$, to plot a 2D-contour of $\ln(\mathcal{L}_{fit})$ over $q$ and $\delta$. This is given in the right panel of Fig.~\ref{fig:fit to WMAP}. We can see that there is a double-peak structure along the $\delta$ axis, separated at about $0.7\pi$. We have confirmed that these two peaks give very similar resulting power spectra. It is not strange to see a double-peak separated by $0.7\pi$, because all large-scale perturbations are more or less spatially periodic.

Although the fitting in Fig.~\ref{fig:fit to WMAP} looks nice, we must be careful about concluding that the entire large-scale CMB asymmetry is generated by our model. At least now, we can only say that part of the large scale power asymmetry can be explained by our toy-model. For example, when we look at the quadrupole ($l=2$) component, we see that the fitting here is not good enough. However, the fitting at $l=2$ can actually be made much better than Fig.~\ref{fig:fit to WMAP}, but at the cost of worse fitting on all other components. Therefore, it's more likely that the quadrupole anomaly is more or less affected by a different origin, and the curvaton scenario is hence not the unique source of all asymmetry/anomaly.

\subsection{Determining the most optimal orientation of the model}
\label{sec:axes}
Since the temperature fluctuations caused by plane wave curvaton perturbations are rotationally symmetric around the wave vector $\vec{k}$ (see Fig.~\ref{fig:example from simulation} for an example), its spherical harmonic components $\alpha_{lm}$ should not have the same strength at different $m$. Especially, if $\vec{k}$ has the same direction to $\pm Z$-axis, then $\alpha_{lm}=0$ for all $m\ne0$ components. Therefore, if the model in this work is real, then the orientation that minimizes the $m\ne0$ spherical components for the real CMB data is very likely the orientation of the model. According to Fig.~\ref{fig:fit to WMAP}, we see that for $l=3\sim7$ components, the parity asymmetry is most significant, thus if we work in this range, it may increase the accuracy of determining the orientation of the model.

\begin{figure}[h]
\begin{center}
  \includegraphics[width=0.62\textwidth,angle=90]{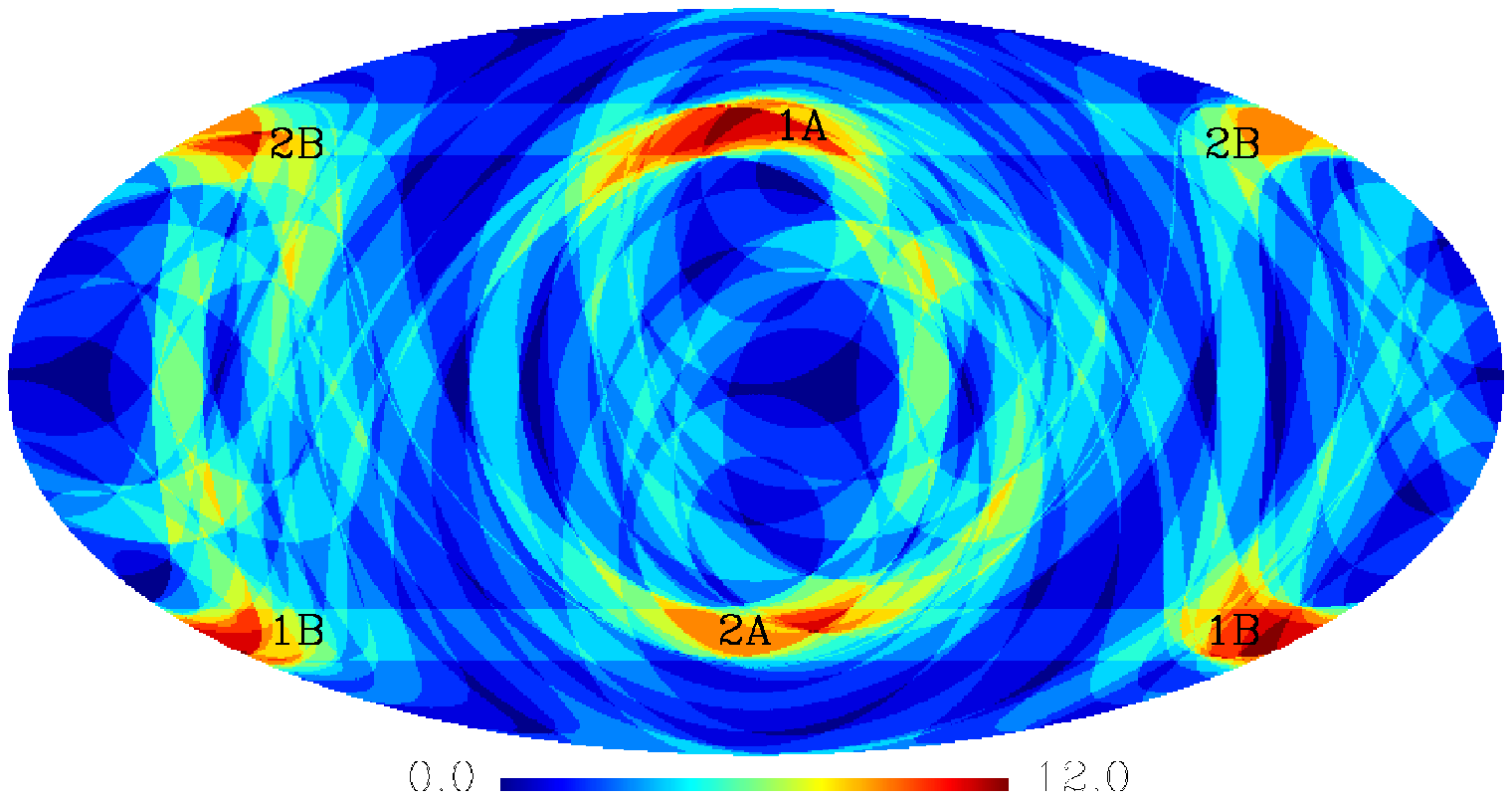}
  \includegraphics[width=0.62\textwidth,angle=90]{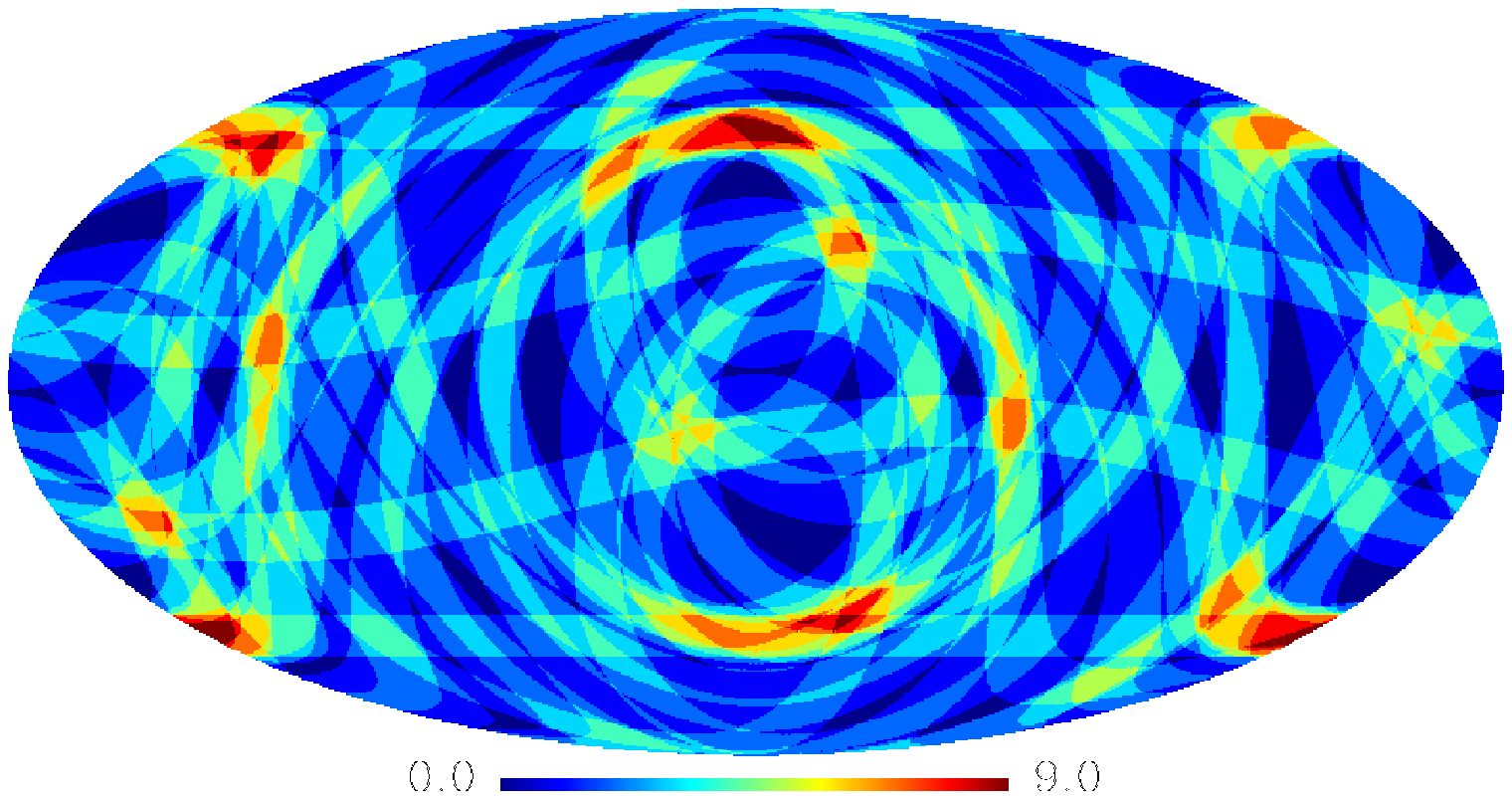}
 \caption{Overlapping of belts corresponding to $|\theta-\pi/2|=\langle\theta_l\rangle$ for $l=3\sim7$, plotted in the Galactic coordinate system. \emph{Upper:} derived from WMAP ILC. \emph{Lower:} derived from Planck NILC. The two poles of the axes with the strongest overlap are marked out by "1A", "1B", and the two poles of the secondary axes are marked out by "2A", "2B".}
\label{fig:belts}
\end{center}
\end{figure}

Our approach is like this: First we rotate the WMAP 7-year ILC map around the $Y$-axis (Galactic plane) to find an angle $\theta_l$ that minimizes $\sum_{m\neq0} |\alpha_{lm}|^2$ for each $l$ in range $l=3\sim7$ respectively. The average value $\langle \theta_l \rangle$ tells us the latitude of the orientation in the Galactic coordinate system, which is $-53^\circ$. Interestingly, as discovered by~\cite{Oliveira04} the preferred axis of the WMAP quadrupole ($l=2$) and octupole ($l=3$) both point to $(l,b)\sim(110^\circ,60^\circ)$ in Virgo. Since the preferred axis does not distinguish between $\hat{n}$ and $-\hat{n}$ ($b=\pm60^\circ$), we see that the axis we have found is only $7^\circ$ away in latitude from the well known "axis-of-evil". Moreover, as discovered by~\cite{Cruz2005MNRAS.356...29C,Cruz2007ApJ...655...11C}, the well known non-Gaussian cold spot at $(l,b)=(209,-57)$ is also only $4^\circ$ away in latitude from our axis here. Thus our model may play an important role not only in the parity asymmetry problem, but also in other well known large-scale CMB anomalies, and perhaps even be connected to some CMB non-Gaussianities.

The standard deviation of $\theta_l$ in range $l=3\sim7$ is $\sigma_0=15.4^\circ$. Such a small value means that these harmonic components have clustered orientations. To confirm that $\sigma_0$ is really small, we did a test with 5000 simulations, and for each $i$-th simulation, the standard deviation of $\theta_l$ in range $l=3\sim7$ was calculated as $\sigma_i$. We only got 41 out of 5000 simulations that had $\sigma_i<\sigma_0$. This means that for the WMAP data the clustering of $\theta_l$ in the range $l=3\sim7$ is significant at a level of $99.2\%$.

After determining the latitude of the orientation, we change the coordinate system of the ILC map by rotating the $Z$-axis to all 192 directions defined by the HEALPix resolution $N_{side}=4$~\citep{gor05}. For each new coordinate system, we do the same as we did above and get $\langle\theta_l\rangle$ for this coordinate system. If the harmonic components are sufficiently clustered (i.e. if $\sigma_0<20^{\circ}$) we draw a belt at $\langle\theta_l\rangle$ with a width of $12^{\circ}$ for this coordinate system. The overlapping of these belts are shown in Fig.~\ref{fig:belts} (all turned to Galactic coordinates). The hottest spots give the possible orientations of the model. The two poles of the strongest orientations are marked out by "1A", "1B". The coordinate of the axis is $(l,b)=(189^{\circ},-55^{\circ})$. We can also see an orientation at $(l,b)=(346^{\circ},-50^{\circ})$ from Fig.~\ref{fig:belts}, whose poles are marked out by "2A", "2B". To determine the most optimal orientation of the model we calculated the correlation coefficients (for $l=3\sim7$) between the multipole components of our model and the ILC, when the model was rotated to directions "1B" (axis-1) and "2A" (axis-2) respectively. As is seen from Table.~\ref{tab:corr} we have determined that axis-1 is the most optimal orientation.

\begin{table}[h]
\center
\begin{tabular}{ccccccc}
  \hline \hline
$l$ 	& 2 		& 3 	& 4 		& 5 		& 6 		& 7 \\ \hline
Axis-1 	& -0.13		& 0.55	& -0.16		& { }0.60	& { }0.59	& -0.05 \\
Axis-2	& { }0.15	& 0.15	& { }0.09	& -0.08		& -0.01		& { }0.14 \\
  \hline \hline
\end{tabular}
\caption{The model-to-ILC correlation coefficients between their multipole components when the model temperature map is rotated to the "1B" (axis-1) and "2A" (axis-2) directions.}
\label{tab:corr}
\end{table}

As also shown in Fig.~\ref{fig:belts}, we have tested the direction calculation using the Planck NILC map~\citep{planck13_overview}, and seen that the result is quite close to WMAP. By also taking into consideration the Planck official results on the low-$l$ anomalies~\citep{planck13_isotropy}, it seems that the directions shown in Fig.~\ref{fig:belts} is not due to systematics, but more likely intrinsic cosmic features or at least residual foreground.
\subsubsection{Similarity between harmonic components}
\label{sub:component similarity}
\begin{figure}[h]
  \includegraphics[width=0.205\textwidth,angle=90]{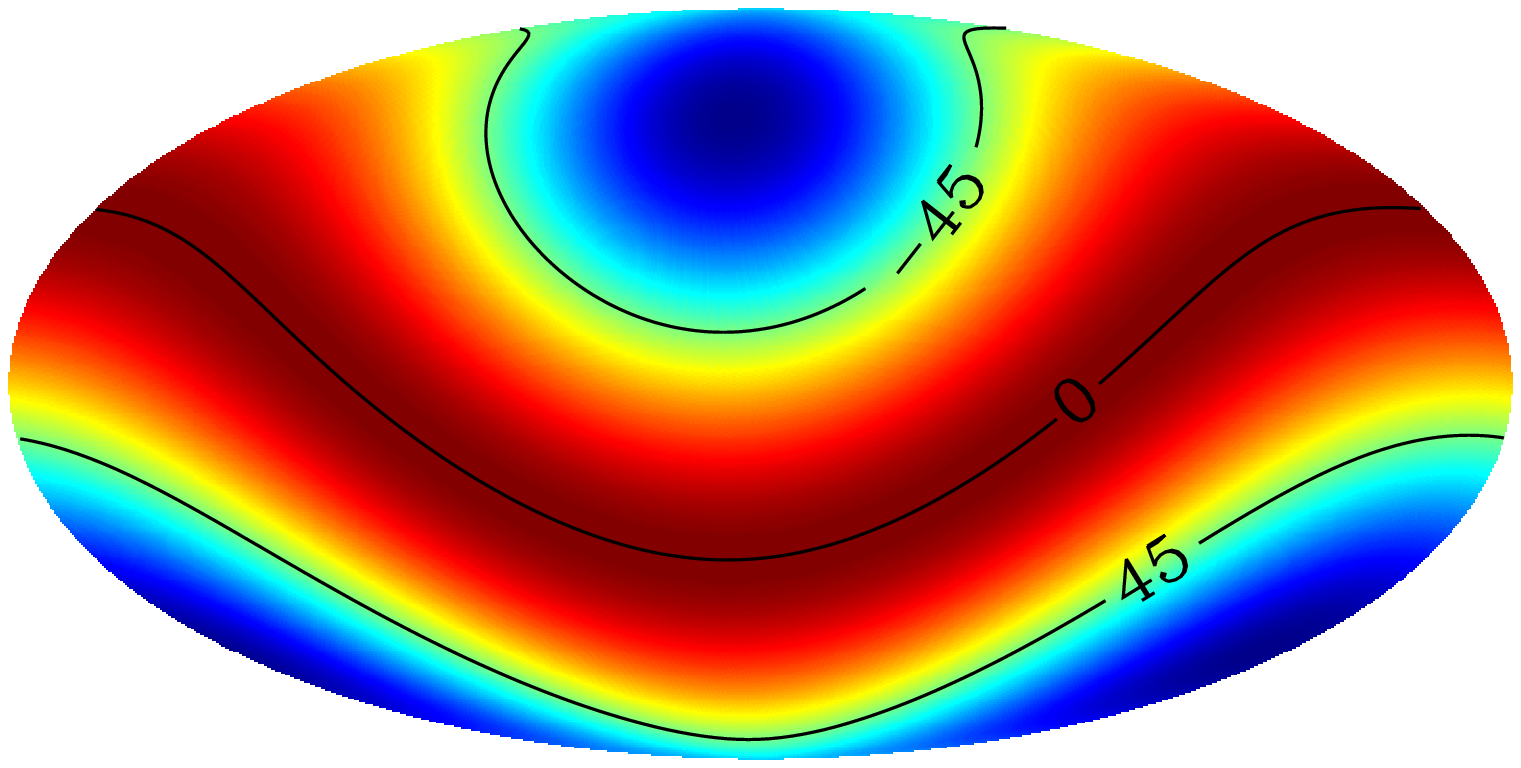}
  \includegraphics[width=0.205\textwidth,angle=90]{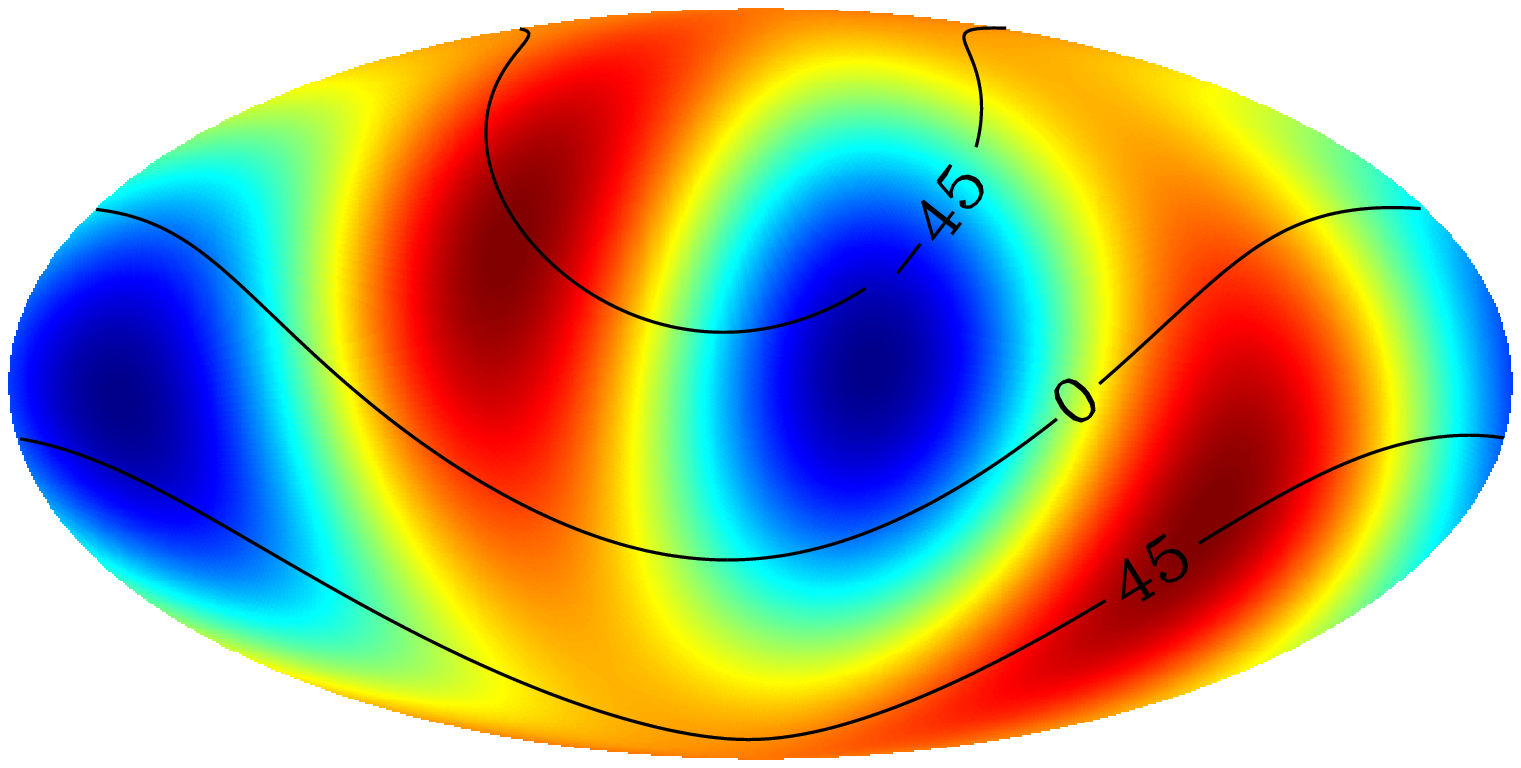}
  \includegraphics[width=0.205\textwidth,angle=90]{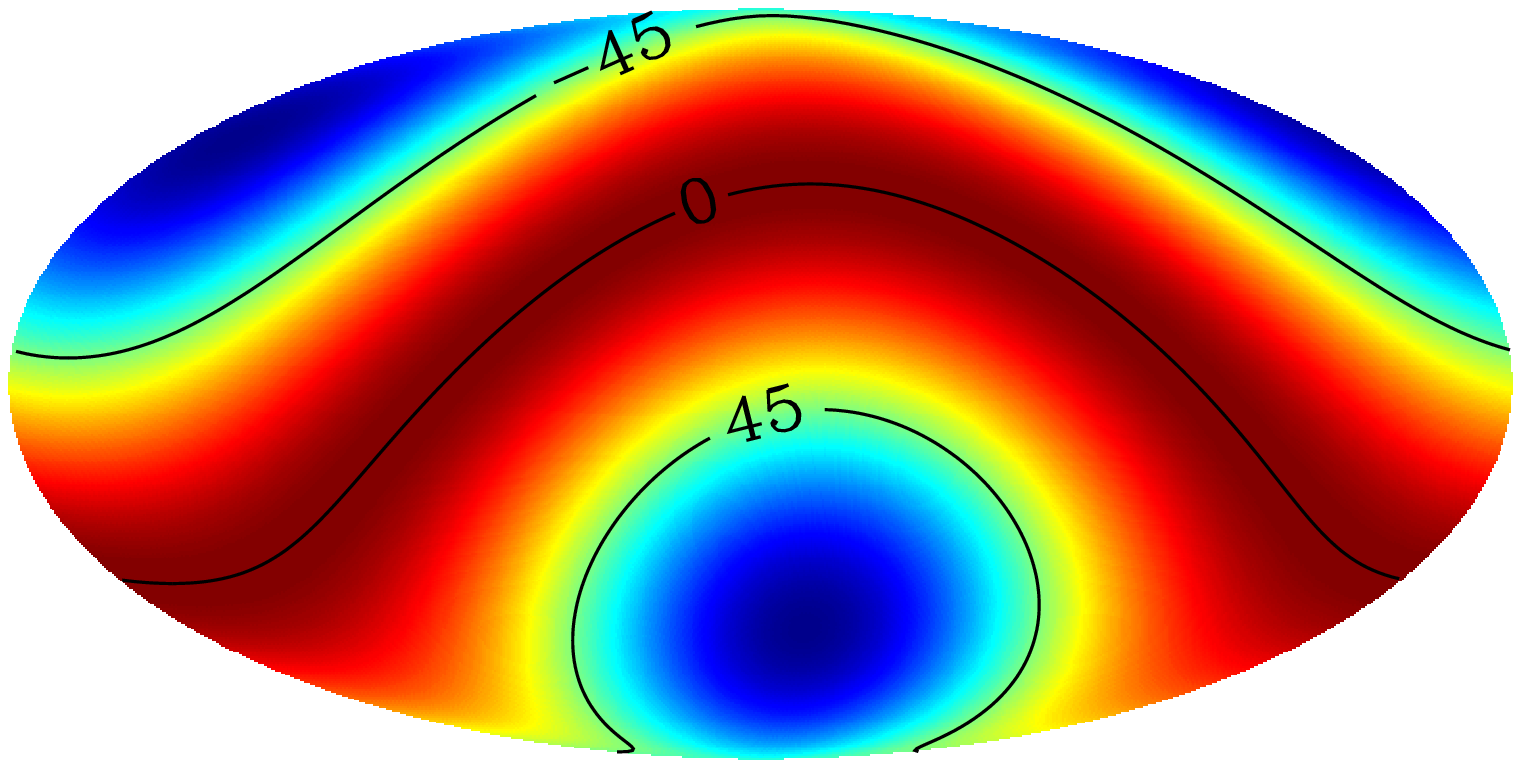}\\
  \includegraphics[width=0.205\textwidth,angle=90]{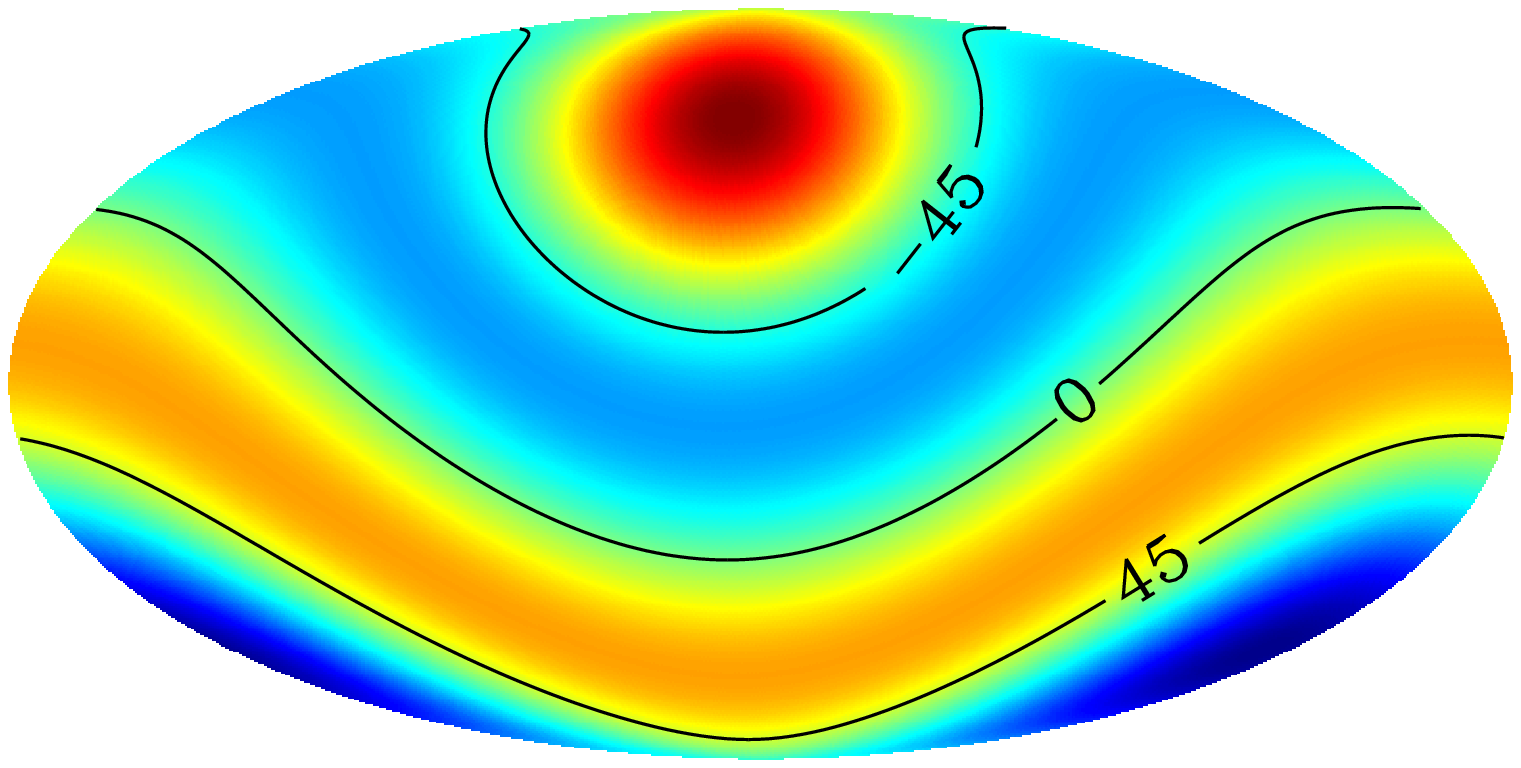}
  \includegraphics[width=0.205\textwidth,angle=90]{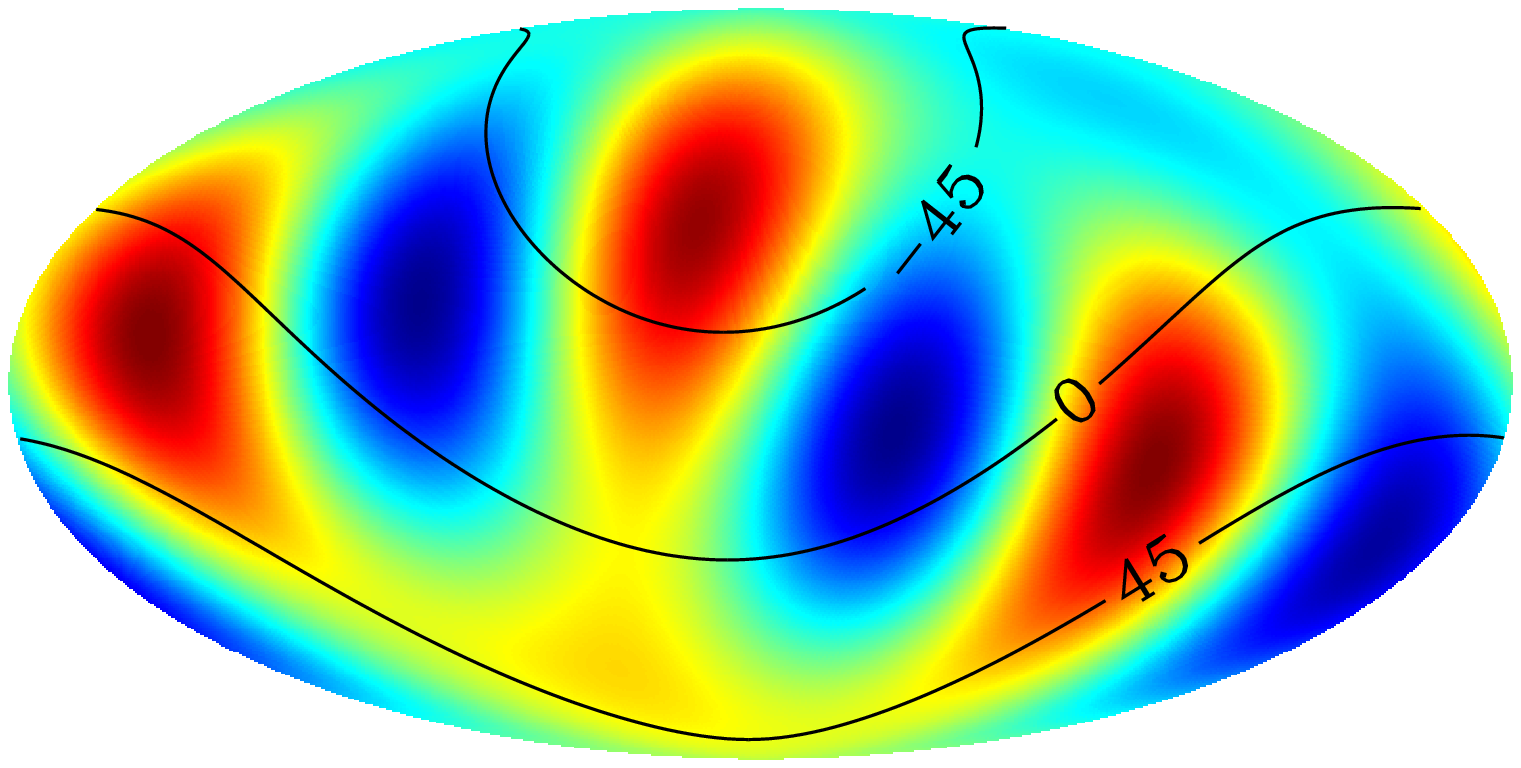}
  \includegraphics[width=0.205\textwidth,angle=90]{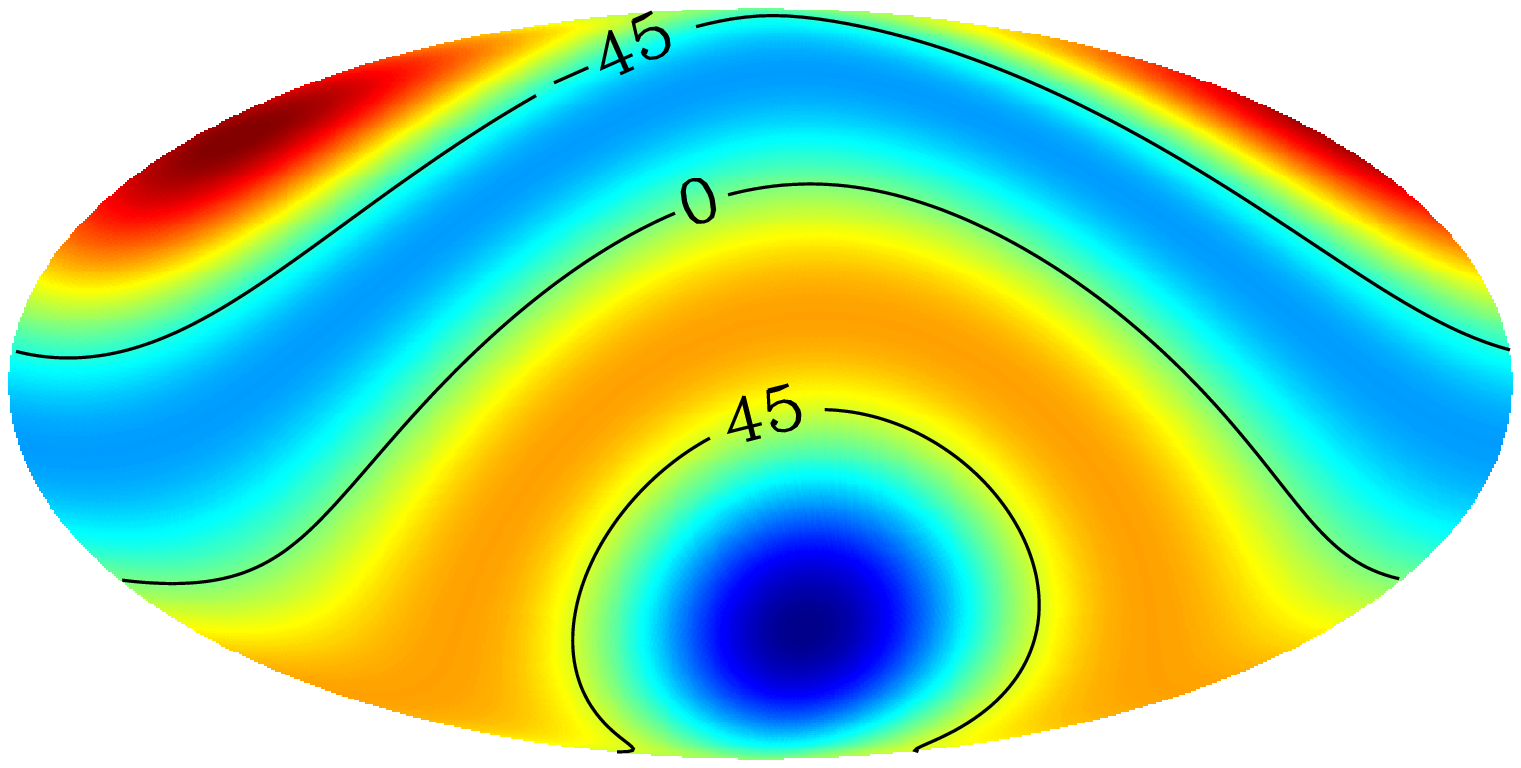}\\
  \includegraphics[width=0.205\textwidth,angle=90]{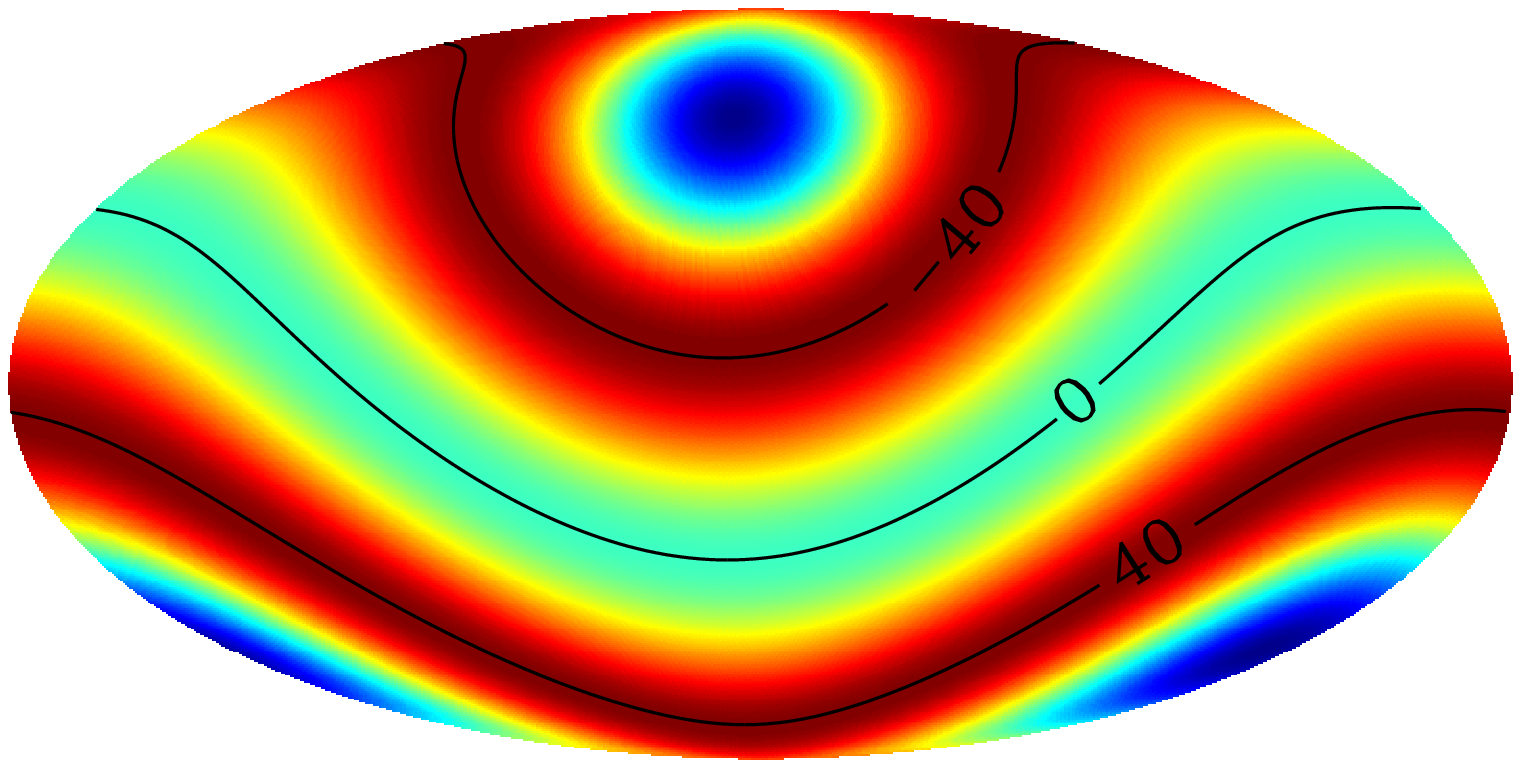}
  \includegraphics[width=0.205\textwidth,angle=90]{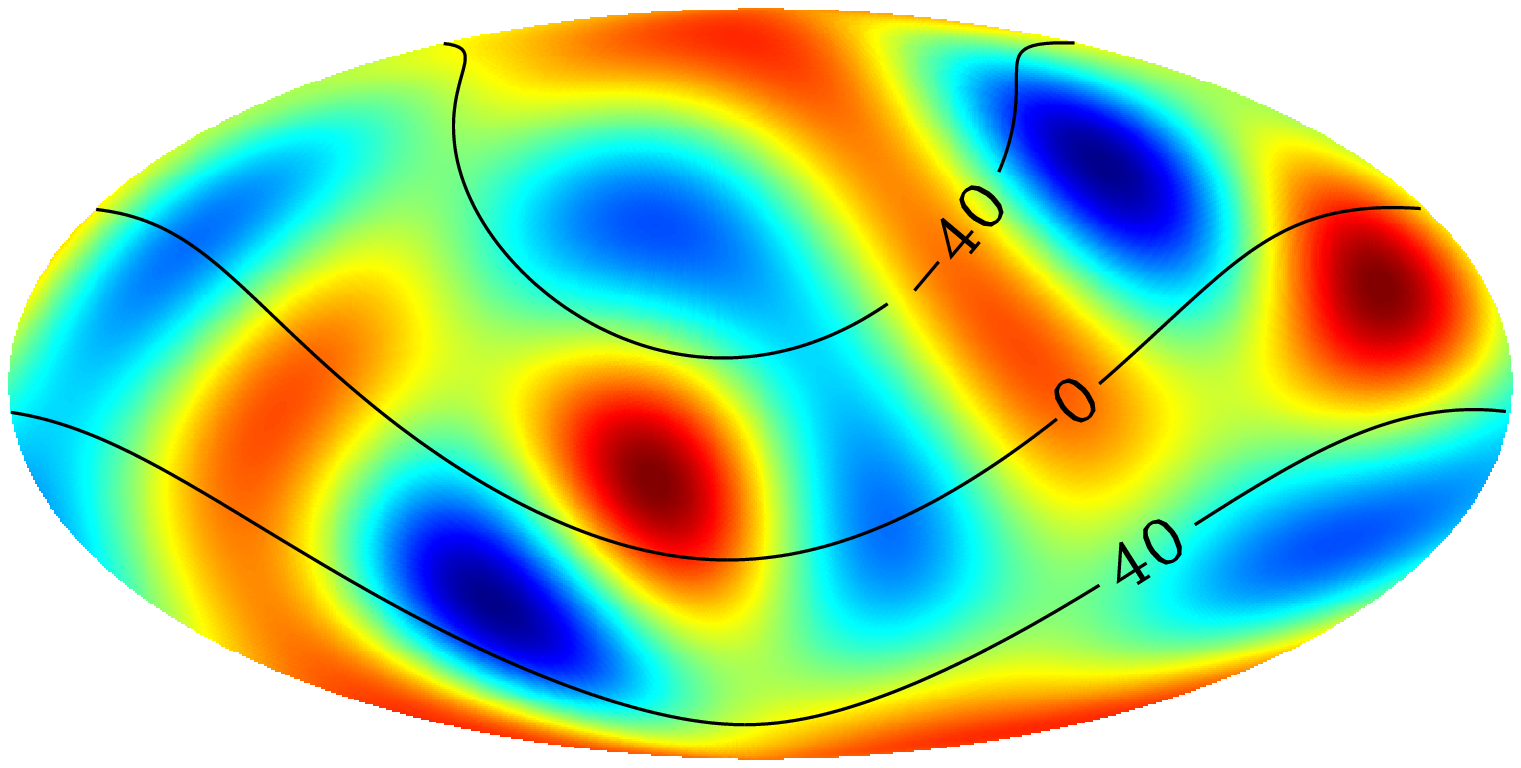}
  \includegraphics[width=0.205\textwidth,angle=90]{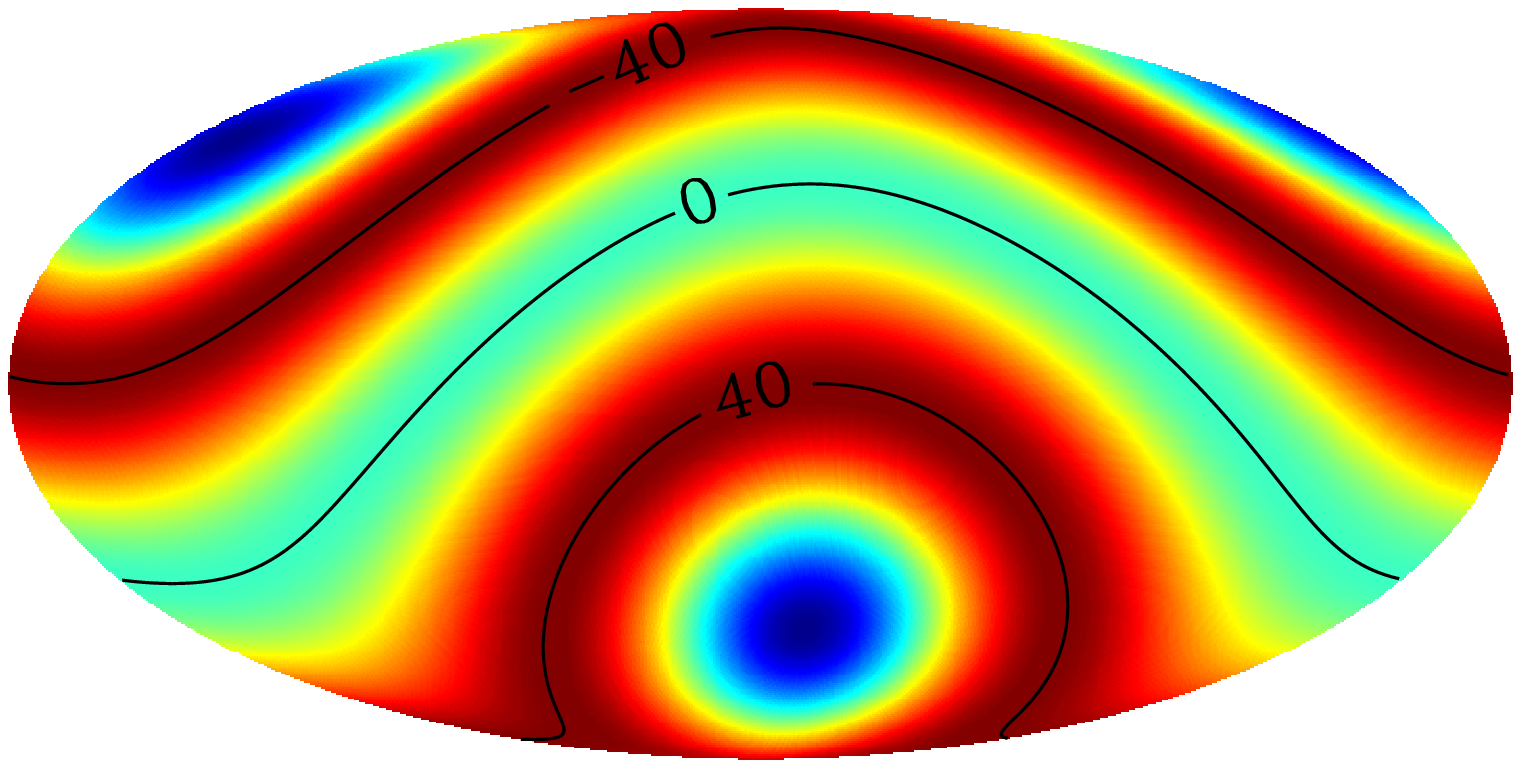}\\
  \includegraphics[width=0.205\textwidth,angle=90]{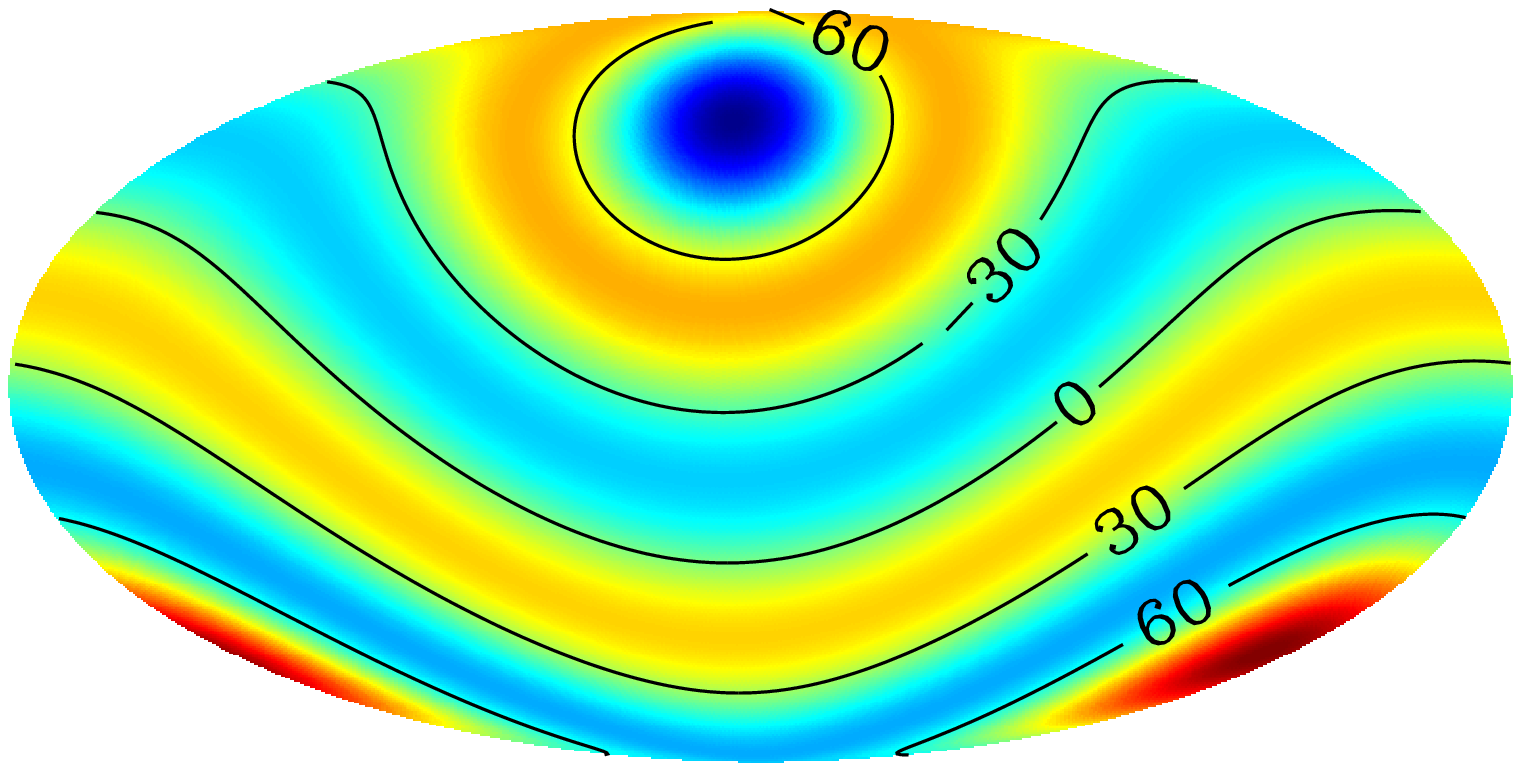}
  \includegraphics[width=0.205\textwidth,angle=90]{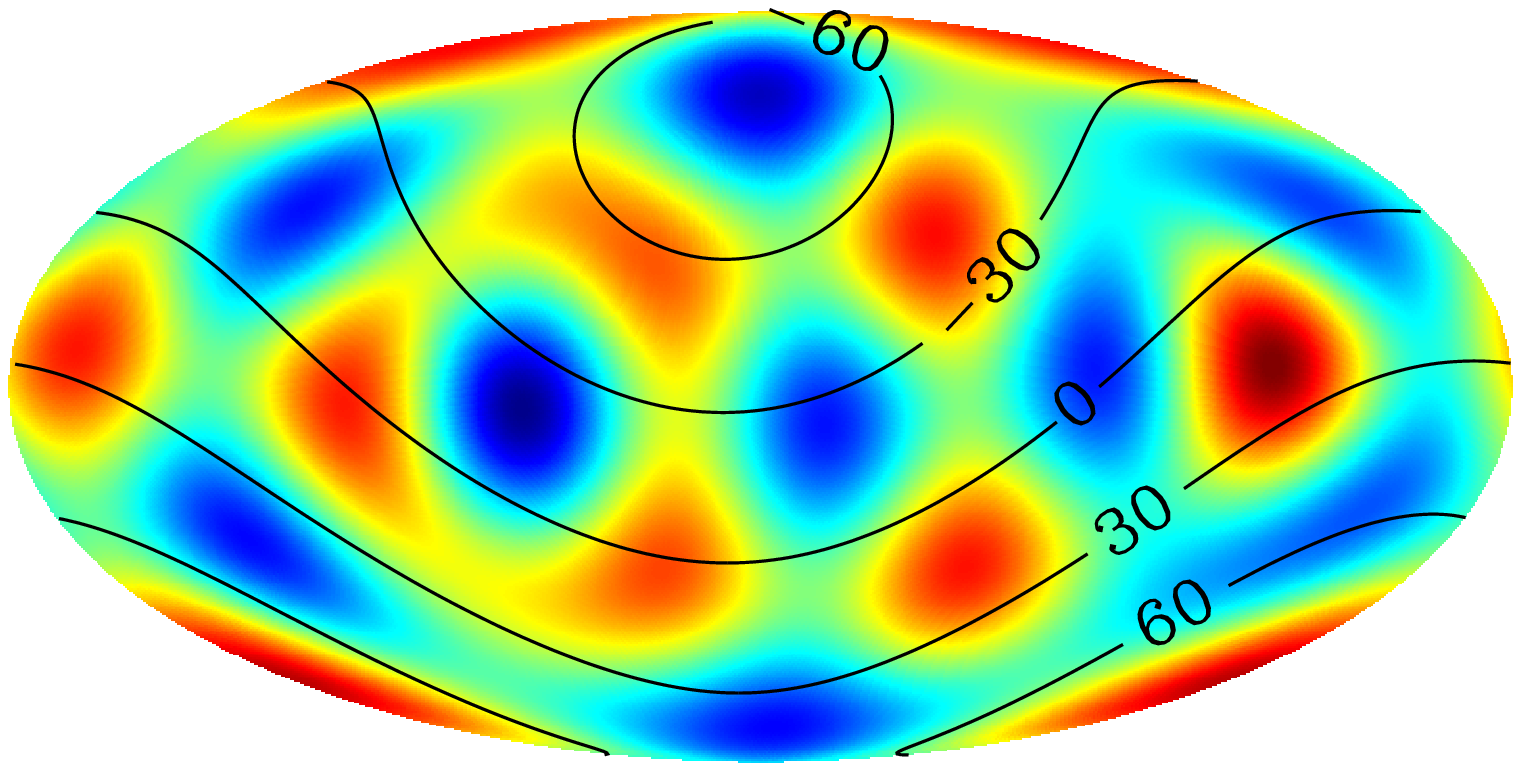}
  \includegraphics[width=0.205\textwidth,angle=90]{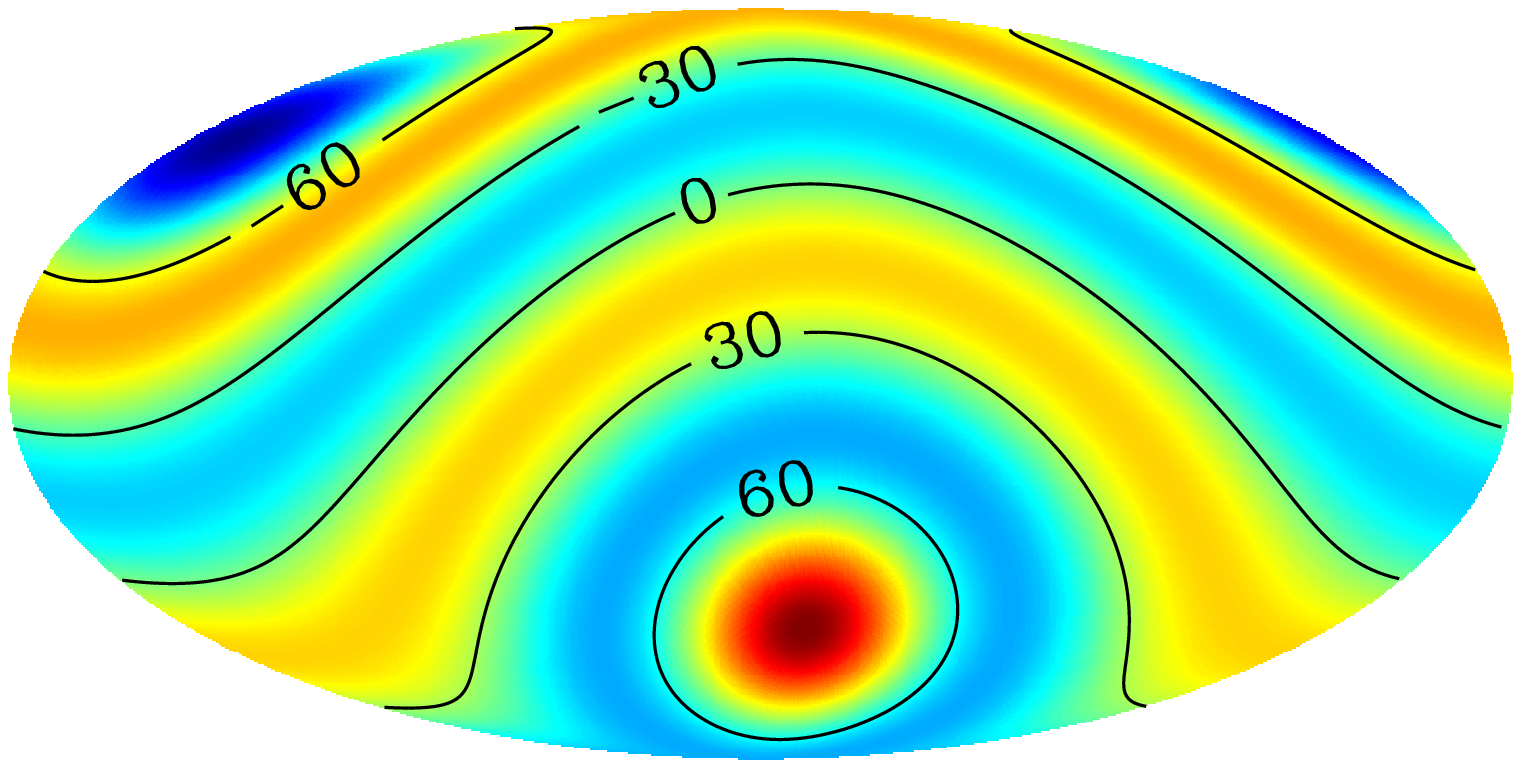}\\
  \includegraphics[width=0.205\textwidth,angle=90]{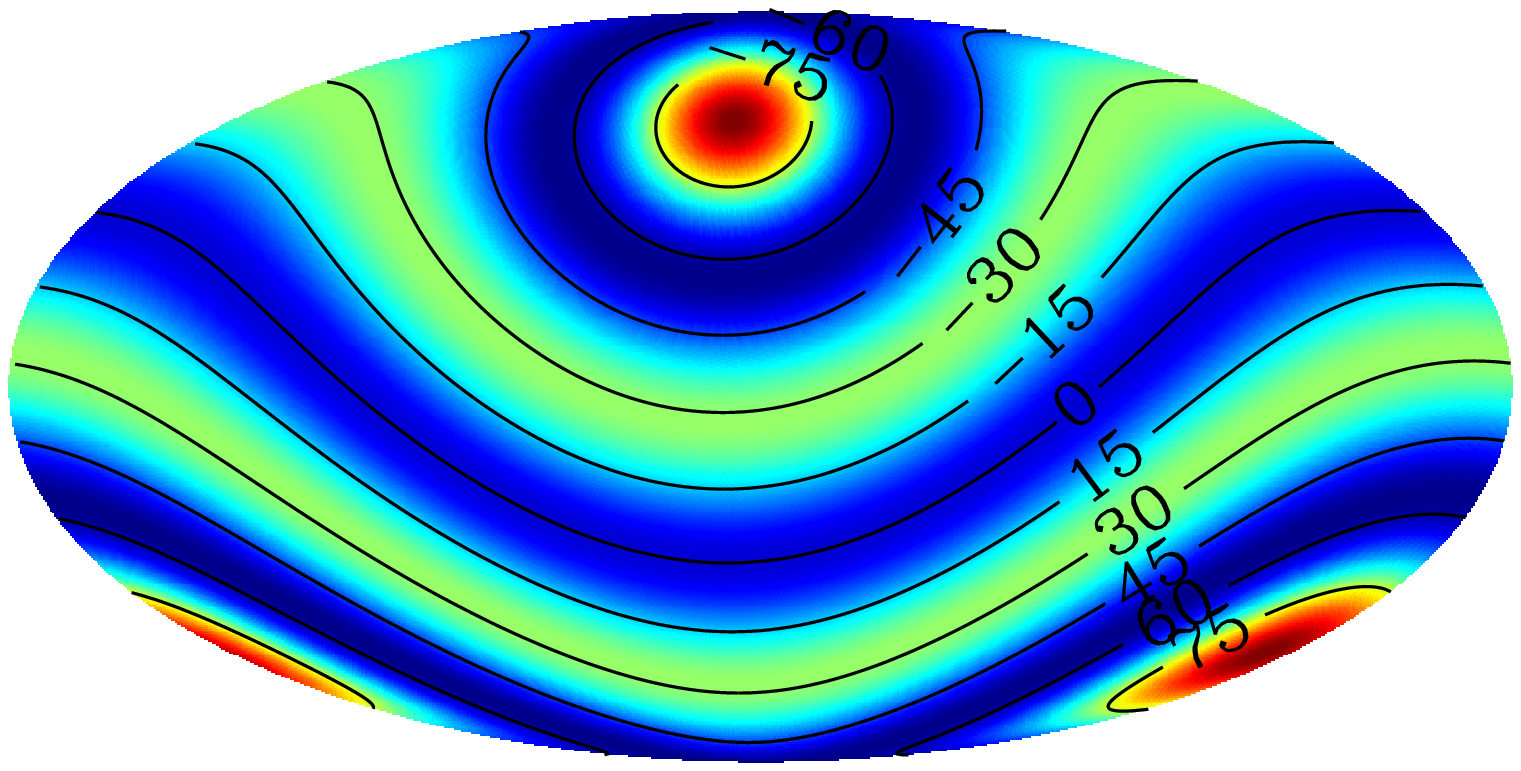}
  \includegraphics[width=0.205\textwidth,angle=90]{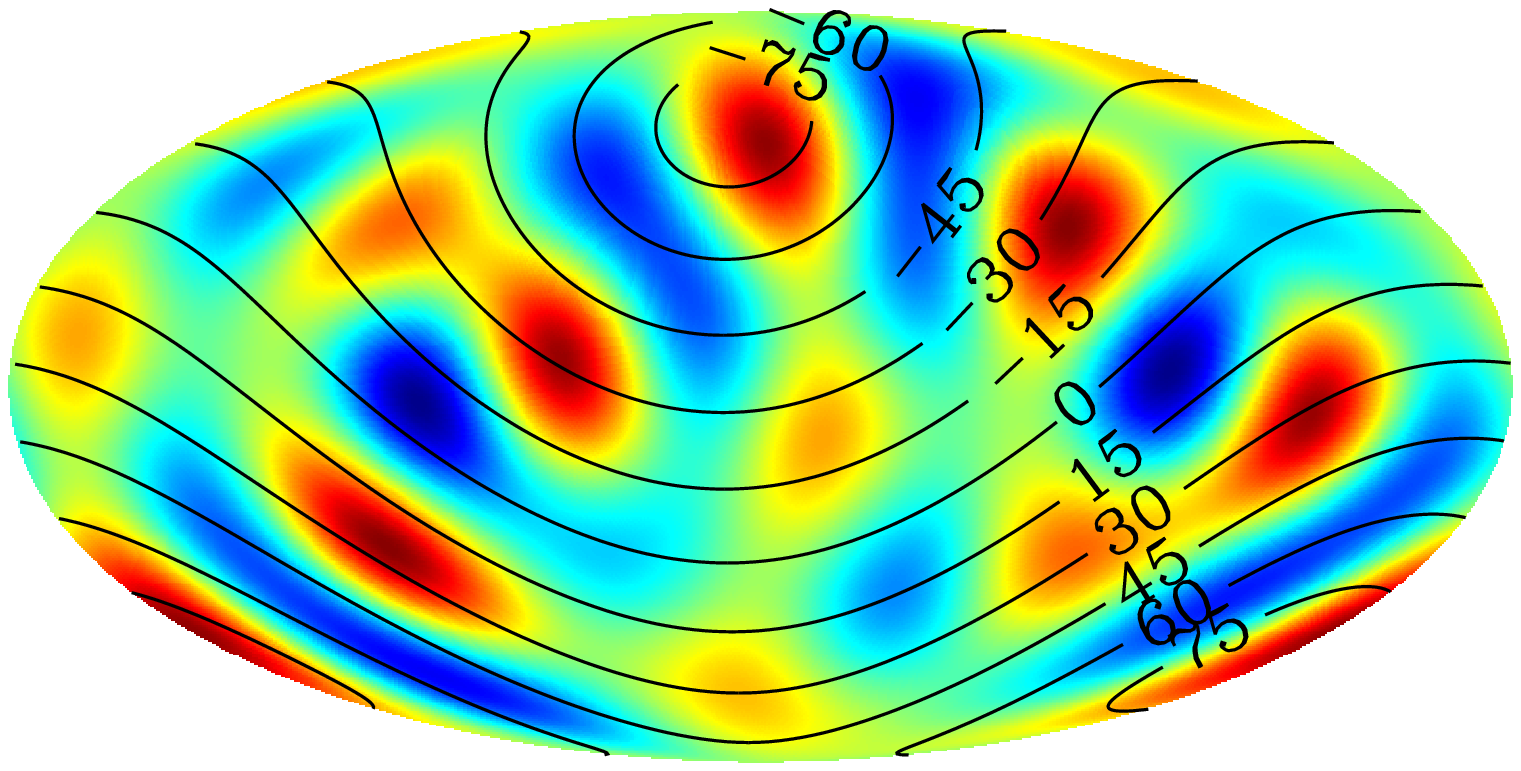}
  \includegraphics[width=0.205\textwidth,angle=90]{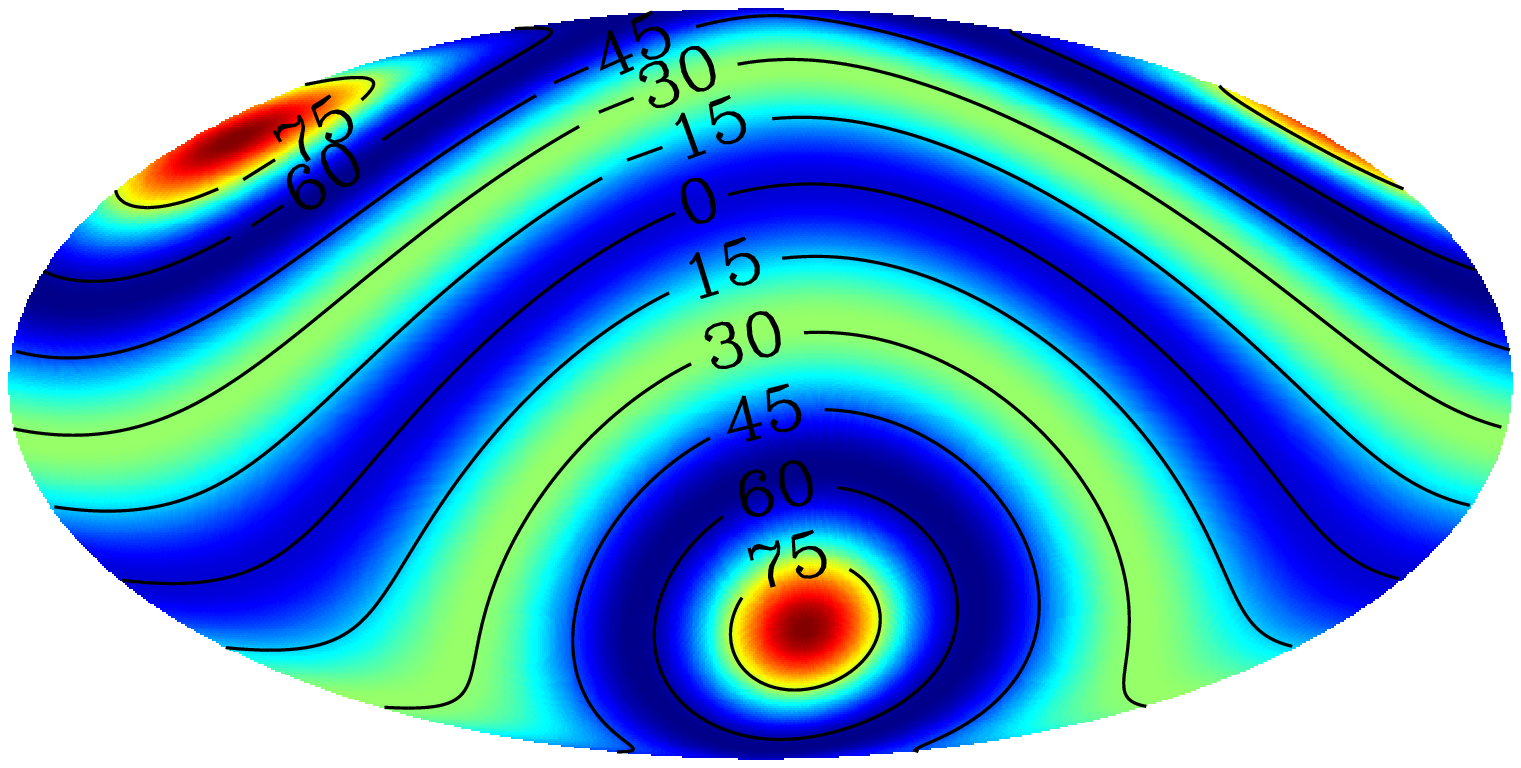}\\
  \includegraphics[width=0.205\textwidth,angle=90]{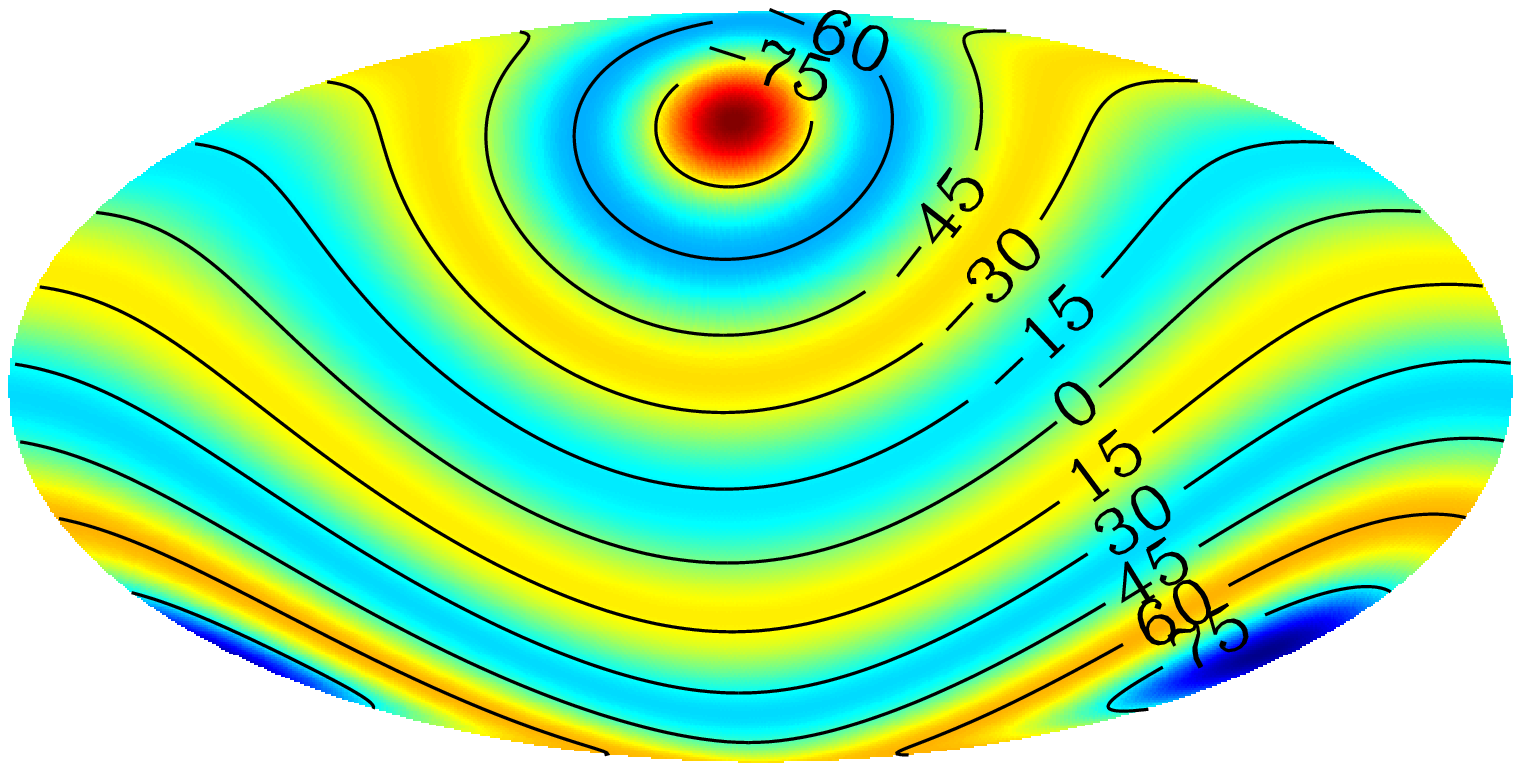}
  \includegraphics[width=0.205\textwidth,angle=90]{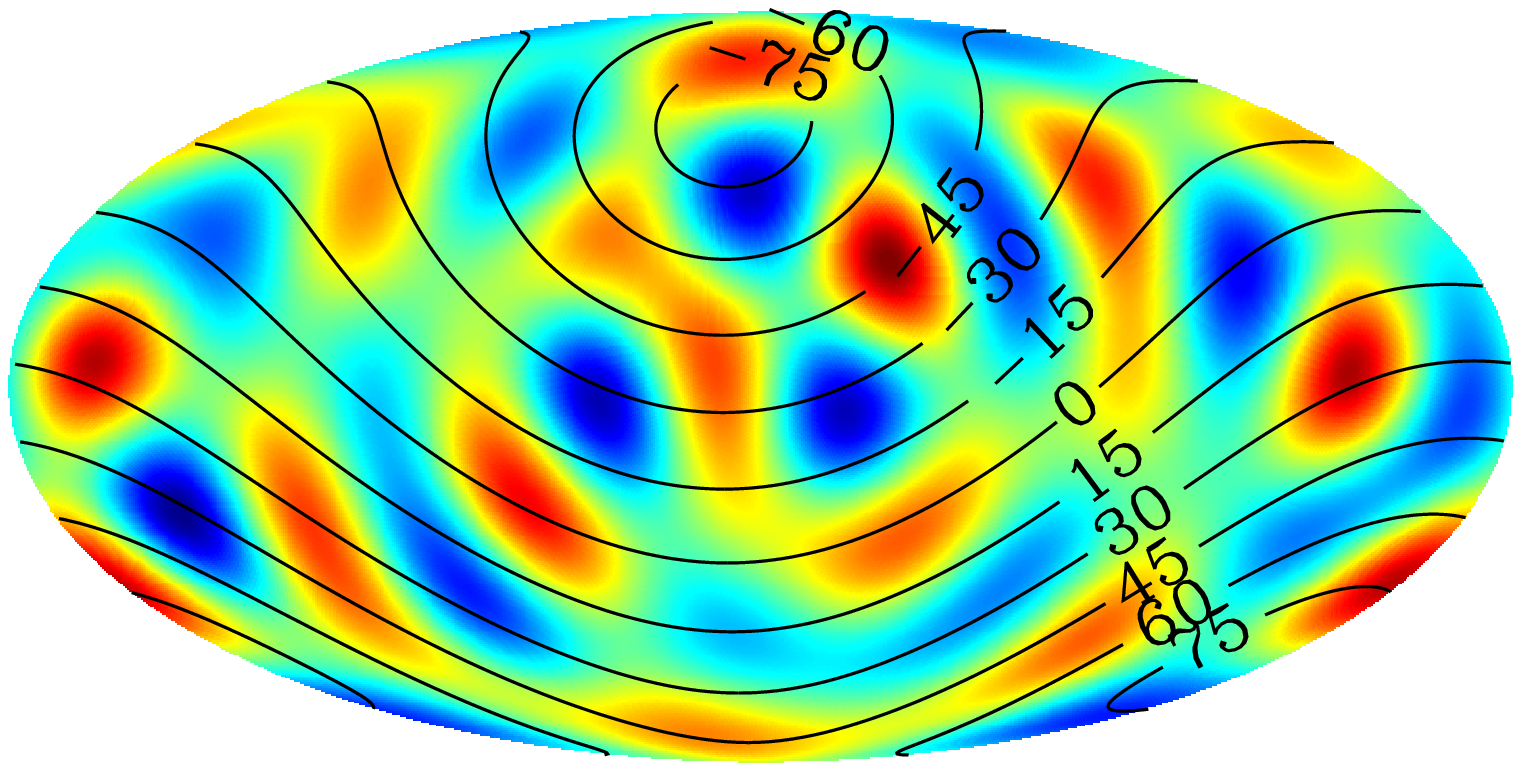}
  \includegraphics[width=0.205\textwidth,angle=90]{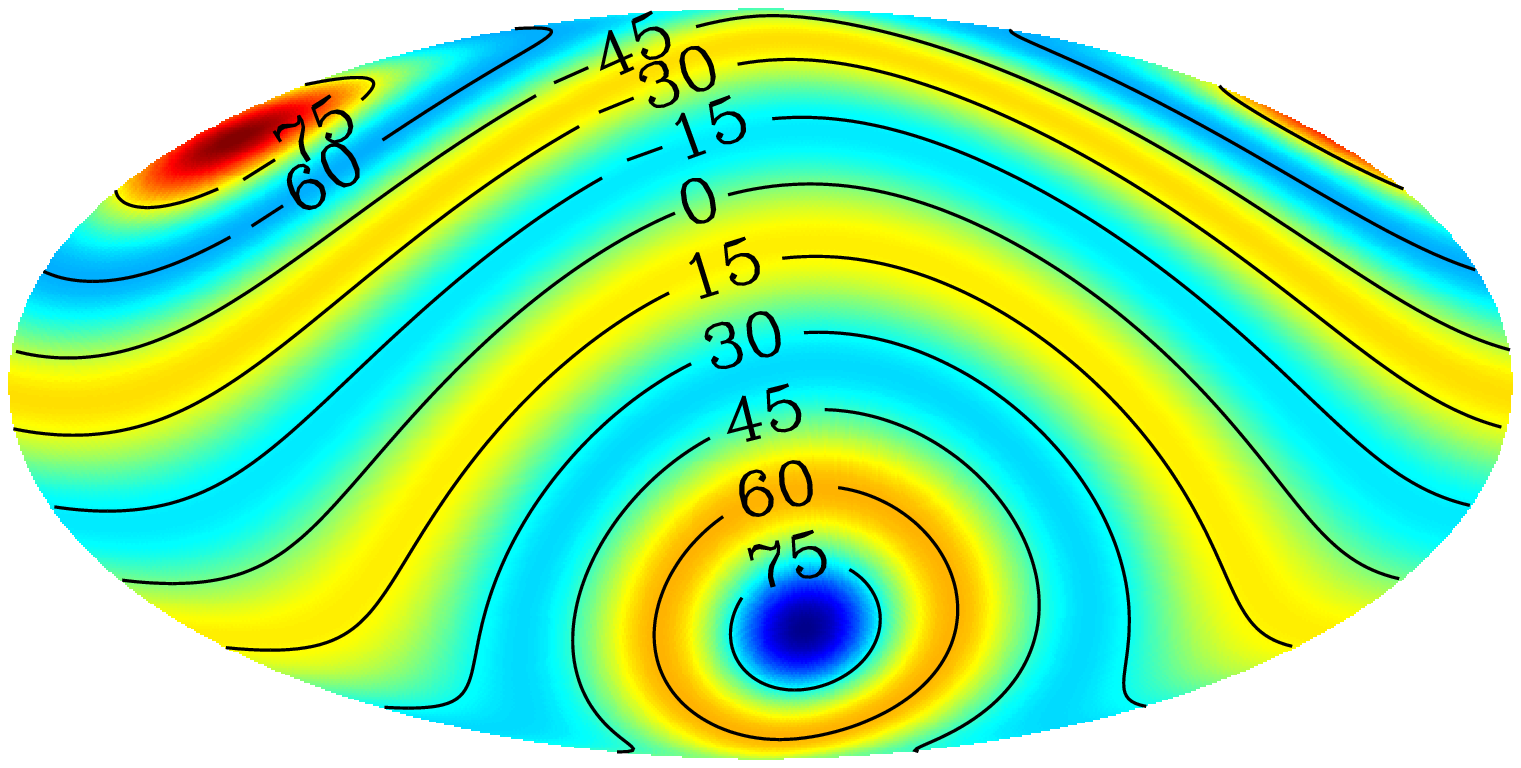}
 \caption{Maps of $l=2\sim7$ (top to bottom). (\emph{Left:}) components of the model with the best-fit parameters shown in Fig.~\ref{fig:fit to WMAP} rotated to the direction of ``1B'' (axis-1). (\emph{Middle:}) the WMAP 7-year ILC map. (\emph{Right:}) the model components rotated to the direction of ``2A'' (axis-2). The lines indicate latitudes in the coordinate system of the model (the lines on the ILC maps correspond to the coordinate system of axis-1).}\label{fig:structure similarity}
\end{figure}

We show the similarity between the large-scales components of the model (rotated to axis-1) and the WMAP ILC map in Fig.~\ref{fig:structure similarity} (the result for Planck NILC is close to this). Let us take the $l=5$ component (fourth row) as an example: both model and the ILC have one cold spot located at $-90^\circ$ to $-60^\circ$. We also see a band of cold spots between $-30^\circ$ and $0^\circ$ and between $30^\circ$ to $60^\circ$. There is one hot spot located at $60^\circ$ to $90^\circ$, and hot spots in the band between $0^\circ$ and $30^\circ$, and the band between $-60^\circ$ and $-30^\circ$. We can conclude that our rotated model fits the structure of the ILC well.

\subsection{Stability of the model to contaminations}\label{sub:test the method for orientation}
We have tested the stability of our method by using an input map derived from Equation~(\ref{equ:CMB due to curvaton}) with $q=3$, $r=0$, $\delta=0$, $\hat{k}=(1,0,0)$ (neglecting the coefficient $4r\Psi_c$). By using the same method mentioned in the previous section, the resulting axis is found to be $(l,b)=(4.5^{\circ},6^{\circ})$, well consistent with expectation, and validates the stability of our method. Then we added two contaminations to test the stability of this method: The contaminations are similar to the source but with $1/10$ strength and different wave vector directions: $\hat{k}=(0,0,1), \sqrt{3}/3(1,1,1)$ respectively. With the same approach, we got a resulting axis at $(b,l)=(10^{\circ},357^{\circ})$, also close to expectation. Thus we have shown that our approach is insensitive to weak contaminations with "misleading" axes (e.g. due to other sources of asymmetry).

\section{The effect on the power spectrum of a plane wave component}
\label{sec:example from simulation}
\begin{figure}
\centering
  \includegraphics[width=0.31\textwidth,angle=90]{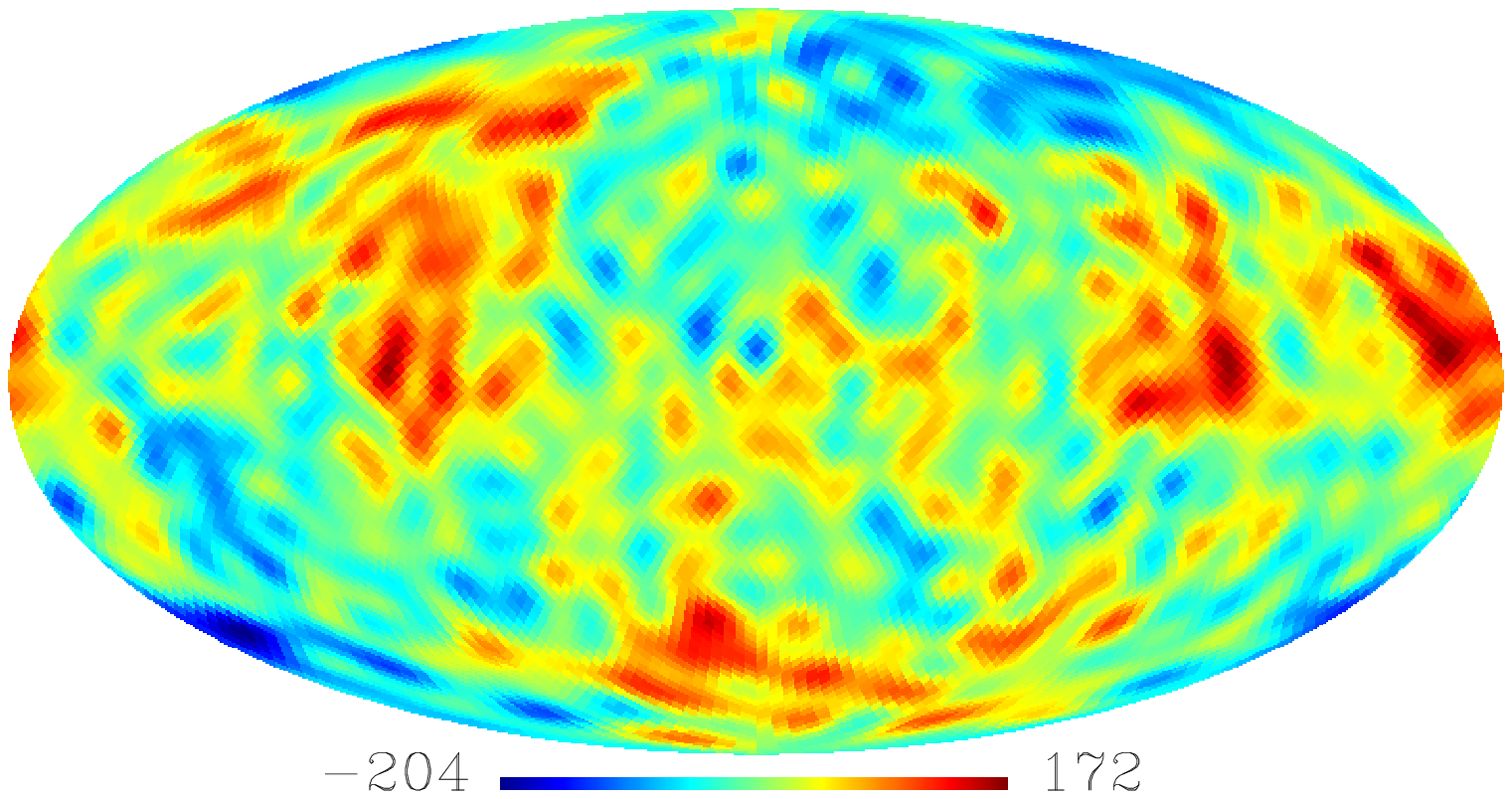}
  \includegraphics[width=0.31\textwidth,angle=90]{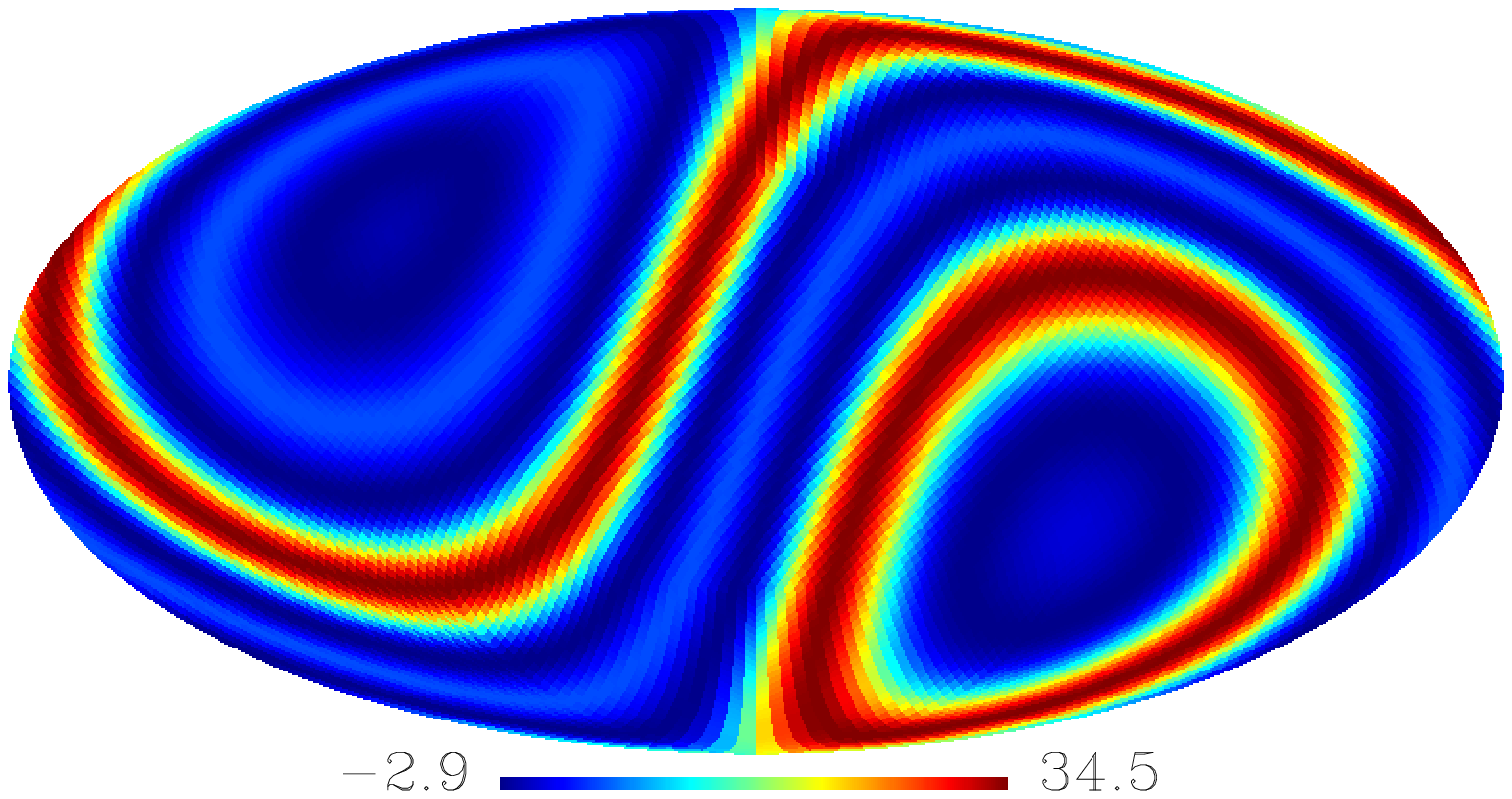}
  \includegraphics[width=0.31\textwidth,angle=90]{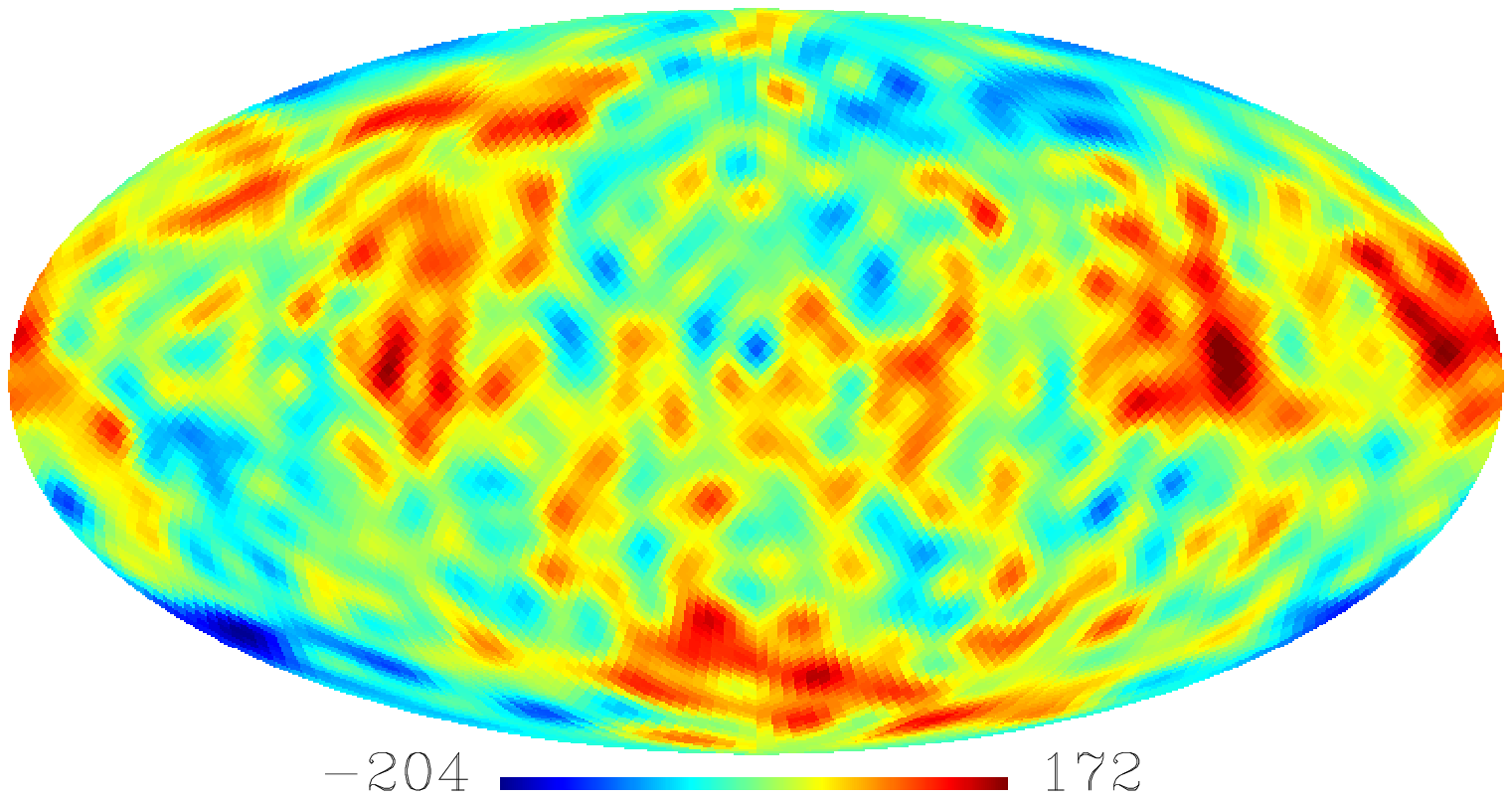}
  \includegraphics[scale=0.43]{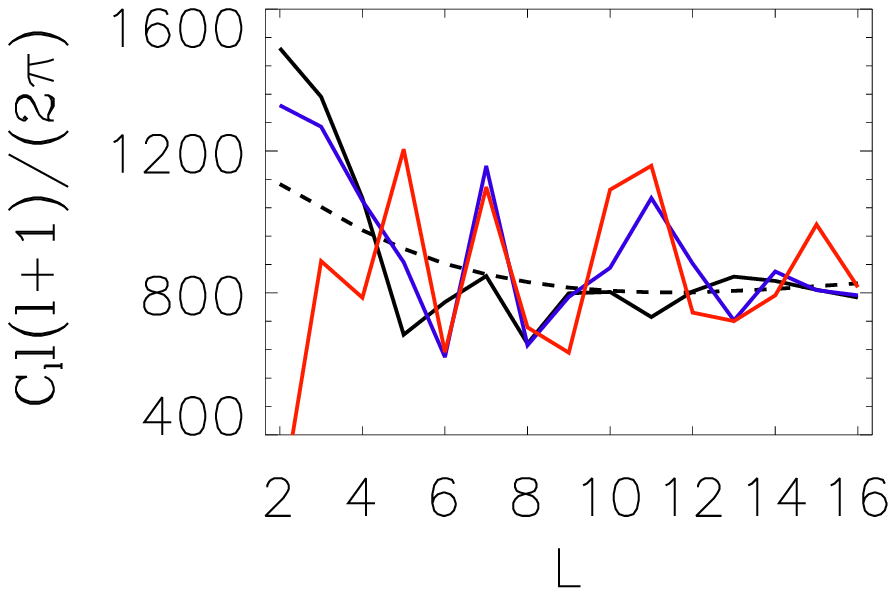}
  \caption{Top left: The original simulated CMB map with $\Lambda$CDM power spectrum. Top right: The plane wave component due to a curvaton field. Bottom Left: The summation of the two top panels. Bottom right: The harmonic power spectrum of: theoretical $\Lambda$CDM (Black dash), simulated map without plane wave (black solid), with plane wave (blue), and WMAP low-$l$ power spectrum (red).}\label{fig:example from simulation}
\end{figure}

Here we show an example from simulation, in which we can clearly see how the power spectrum
power asymmetry can be generated from a plane wave component due to the curvaton model
discussed above. First we generated a simulated CMB map from a $\Lambda$CDM power spectrum. Then we generated a map of a plane wave component
with the same parameters as shown in Fig.~\ref{fig:fit to WMAP} (the direction of the plane wave is not important for a simulated map,
so we choose the Ecliptic north/south poles as the direction). The summation of them resembles
the "$\Lambda$CDM + curvaton" scenario. We then calculate the power spectrum of the original map, do the summation and calculate the total power spectrum,
and plot them together with the WMAP CMB low-$l$ power spectrum in Fig.~\ref{fig:example from simulation}.
We can see that the odd-even multipole power asymmetry of the WMAP low-$l$ power spectrum can be very well reproduced in this way (with the exception of the quadrupole ($l=2$)).
We have thus shown that the model presented in this work can reasonably account for the WMAP low-$l$ power asymmetry.

Certainly, with more simulations we can also see cases in which the combined power spectrum (blue line in Fig.~\ref{fig:example from simulation}) isn't similar to the real data (red line in Fig.~\ref{fig:example from simulation}). The reason is simple: the CMB and curvaton components can have different directions and phases, thus the summation of them can make the power spectrum either higher or lower. This fact makes the problem much more complex. However, with more simulations we can see that with the curvaton component presented in Fig.~\ref{fig:fit to WMAP}, the probability of getting similar result to real data will increase.

The power spectrum similarity is evaluated by the cross correlation coefficient between the power spectrum for WMAP data and simulations in the range $l=4\sim12$ for 10,000 simulations:
\begin{equation}\label{equ:correlation}
C_{4-12}=Corr(C_l^{sim},C_l^{WMAP}), \quad \quad l=4\sim12.
\end{equation}
If the simulations are pure $\Lambda$CDM (no curvaton component is added), only $2.6\%\pm0.16$ of the simulations have a $C_{4-12}>0.6$. For the simulations with $\Lambda$CDM + curvaton component $18.4\%\pm0.43$ of the simulations have a $C_{4-12}>0.6$. This fact supports the curvaton scenario presented in this work quite well.

\section{Discussions}
\label{sec:conclusion}
In this work we introduce a model based on the curvaton scenario, which has only three parameters, and try to apply it to the WMAP data. This model is an extension of~\cite{Erickcek.0808.1570,Erickcek.0907.0705}, and the main difference is: We have discovered that if the wave length of the curvaton perturbation is comparable to or smaller than the horizon ($q\geq 1$), then the model can be used to explain parity asymmetry and probably more asymmetry problems. Our results show that such a simple model can give a very well fit to the CMB power spectrum difference between $\Lambda$CDM expectation and experimental detection. The spatial structure can also be well fitted, especially for $l=5$. This tells us that at least part of the CMB large-scale asymmetry can be attributed to an extra component to the inflation field, which provides a possible cosmic explanation to the CMB asymmetry problems. However, these morphological features could also be mimicked by some combination of foreground residuals, as discussed in \cite{Hansen2012}. In this work our goal has been to propose a theoretical model based on the curvaton scenario, and to see which constraints we have to apply to it in order to explain the observed parity asymmetry in the CMB.
It's also interesting to wait for the Planck polarization data to see what will finally happen to TE, EE, EB, as discussed by~\cite{Gruppuso13}.

A recent paper has placed limits on the semi-classical fluctuations in the primordial Universe~\citep{Aslanyan13}. Although the fluctuation amplitude of our model is not discussed here, we notice that they have obtained nearly same direction to us in their Fig. 3 (compare to the "1A" direction in our Fig.~\ref{fig:belts}). Therefore, the limits on fluctuation amplitude discussed in their paper may also apply to our model.

\acknowledgments
We acknowledge the use of the NASA Legacy Archive for extracting the WMAP data. We also acknowledge the use of HEALpix~\citep{gor05}\footnote{\tt http://www.eso.org/science/healpix/}. This work is supported in part by Danmarks Grundforskningsfond which allowed the establishment of the Danish Discovery Center, FNU grant 10-083918, the National Natural Science Foundation of China (Grant No. 11033003), the National Natural Science Foundation for Young Scientists of China (Grant No. 11203024) and the Youth Innovation Promotion Association, CAS. The anonymous referee is thanked for the constructive suggestions.



\end{document}